\documentclass{article}

\usepackage{authblk}
\usepackage[utf8]{inputenc} % allow utf-8 input
\usepackage[T1]{fontenc}    % use 8-bit T1 fonts
\usepackage{hyperref}       % hyperlinks
\usepackage{url}            % simple URL typesetting
\usepackage{booktabs}       % professional-quality tables
\usepackage{amsfonts}       % blackboard math symbols
\usepackage{nicefrac}       % compact symbols for 1/2, etc.
\usepackage{microtype}      % microtypography
\usepackage{lipsum}		% Can be removed after putting your text content
\usepackage{graphicx}
\usepackage{natbib}
\usepackage{doi}
\usepackage{amsmath, amsthm}
\usepackage{booktabs,xcolor,enumitem,tikz}
\usetikzlibrary{shapes.geometric, arrows.meta, positioning, calc}
\hypersetup{colorlinks, citecolor={blue!80!black}, linkcolor={blue!80!black}, urlcolor={blue!80!black}}

\usepackage[verbose=true,letterpaper]{geometry}
\AtBeginDocument{
  \newgeometry{
    textheight=9in,
    textwidth=6.5in,
    top=1in,
    headheight=14pt,
    headsep=25pt,
    footskip=30pt
  }
}
\usepackage{setspace}
\theoremstyle{plain}

\newtheorem{theorem}{Theorem}
\newtheorem{lemma}[theorem]{Lemma}

\newtheorem{proposition}{Proposition}
\newtheorem{corollary}{Corollary}
\newtheorem{algorithm}{Algorithm}

\DeclareMathOperator\bE{\mathbb E} % expectation
\DeclareMathOperator\bV{\mathbb V} % variance
\DeclareMathOperator{\Cov}{\mathrm Cov}

\newcommand{\ATE}{\mbox{\tiny{ATE}}}
\newcommand{\OW}{\mbox{\tiny{OW}}}

\DeclareMathOperator\expit{\mathrm expit}
\DeclareMathOperator\logit{\mathrm logit}

\newcommand{\rev}[1]{{\textcolor{black}{#1}}}

\title{Sample size and power calculations for causal inference in observational studies}

\author[1]{Bo Liu}
\author[2]{Chengxin Yang}
\author[1]{Fan Li}
\affil[1]{Department of Statistical Science, Duke University}
\affil[2]{Department of Biostatistics and Bioinformatics, Duke University}

\begin{document}
\maketitle

\begin{abstract}
This paper investigates the theoretical foundation and develops analytical formulas for sample size and power calculations for causal inference with observational data. By analyzing the variance of an inverse probability weighting estimator of the average treatment effect, we decompose the power calculation into three components: propensity score distribution, potential outcome distribution, and their correlation. We show that to determine the minimal sample size of an observational study, in addition to the standard inputs in the power calculation of randomized trials, it is sufficient to have two parameters, which quantify the strength of the confounder-treatment and the confounder-outcome association, respectively. For the former, we propose using the Bhattacharyya coefficient, which measures the covariate overlap and, together with the treatment proportion, leads to a uniquely identifiable and easily computable propensity score distribution. For the latter, we propose a sensitivity parameter bounded by the R-squared statistic of the regression of the outcome on covariates. Our procedure \rev{relies on a parametric propensity score model and a semiparametric restricted mean outcome model}, but does not require distributional assumptions on the multivariate covariates.
We develop an associated R package \texttt{PSpower}.
\end{abstract}


\section{Introduction}
Randomized controlled trials are the gold standard for causal inference. Designing randomized trials requires sample size and power calculations, in which investigators pre-specify a target estimand---typically the effect size of a treatment---and characteristics of the sample of the intended study. The goal is to determine the minimal sample size needed to achieve the statistical power ($\beta$), given a type I error rate ($\alpha$) level, to test the treatment effect estimated by a specific method based on a sample that satisfies these characteristics, or conversely, to determine the power to detect a treatment effect at the $\alpha$ level given a fixed sample size. Sample size and power calculations are the two sides of the same coin, and we will use them interchangeably henceforth. Power calculation is conducted in the design stage of a study, prior to data collection. Thus, individual data are usually unavailable; investigators at best have some summary information from previous similar studies, e.g. effect sizes. Analysts must rely on this limited summary information to approximate the full data structure. The power calculation for randomized trials is largely simplified due to randomization and has been well established \citep{chow2017sample,branson2024power}.   

When randomized trials are not feasible due to logistical or ethical constraints, researchers are increasingly using observational data to emulate randomized trials. The power analysis for observational studies is challenging and has been less developed. In randomized trials, there is no confounding due to randomization; therefore, one can simply use the difference in mean outcomes between arms to estimate treatment effects. Closed-form formulas, e.g., based on the two-sample z-test, are readily applicable for sample size calculation. In contrast, in observational studies, confounding must be adjusted for to obtain an unbiased estimate of treatment effects. Applying power formulas designed for randomized trials to observational studies often leads to severely underpowered studies. In practice, a general power analysis method is to simulate individual data based on pre-specified summary statistics \citep{mukherjee2003estimating,qin2024sample}. This method usually requires unverifiable assumptions of the entire data generating process and lacks a theoretical basis on what summary statistics are sufficient or necessary. In fact, many different data generating models may satisfy the same summary statistics but lead to vastly different sample sizes; choosing between these models is usually \emph{ad hoc}. Another method is to use the variance inflation factor \citep{hsieh2000sample, shook2022power} to approximate the inflation in the sample size of an observational study compared to a randomized trial.  For example, \cite{shook2022power} derived a variance inflation factor formula based on the analysis of the variance of an inverse probability weighting estimator of the average treatment effect; however, their formula requires knowledge on the distribution of the weights, which is usually not available.

In summary, existing power calculation methods usually require information on the data generating process or the distributions of the weights or covariates, which is not always available in the design stage of a study. In this paper, we bypass this limitation by reducing the variance formula to a function of a few interpretable and commonly available summary parameters instead of the weights or covariates.  Based on the analysis of variance of the H\'ajek inverse probability weighting estimator, we decompose the power calculation into three components: propensity score distribution, potential outcome distribution, and their correlation. We show that to determine the minimal sample size of an observational study, in addition to the standard inputs in the power calculation of a randomized trial, it is sufficient to have two parameters, which quantify the strength of the confounder-treatment and the confounder-outcome association, respectively. For the former, we propose using the Bhattacharyya coefficient, which measures the covariate overlap and, together with the treatment proportion, leads to a uniquely identifiable and easily computable propensity score distribution. For the latter, we propose a sensitivity parameter bounded by the R-squared statistic of the regression of the outcome on covariates. {Our derivation relies on a parametric logistic probability score model and a semiparametric restricted mean outcome model with homogeneous treatment effect.} However, by appealing to the Lyapunov-type Central Limit Theorem, we bypass specific distributional assumptions on the multivariate covariates. We derive analytical formulas for three common causal estimands: the average treatment effect for the overall (ATE), treated (ATT), and overlapped population (ATO). Simulations suggest that our formula is robust under treatment heterogeneity. We develop an associated R package \texttt{PSpower}, available at \url{https://CRAN.R-project.org/package=PSpower}, and an online calculator, available at \url{https://www2.stat.duke.edu/~fl35/pspower.html}.

\section{Variance and sample size based on the H\'{a}jek estimator} \label{sec:variance}
Suppose we have a sample of $N$ units. Each unit $i$ ($i=1,2,\ldots,N$) has a binary treatment indicator $Z_{i}$, with $Z_{i}=0$ being the control and $Z_{i}=1$ being treated, and a vector of $p$ covariates $X_{i}=(X_{1i},\cdots, X_{pi})'$. For each unit $i$, there are two potential outcomes $\{Y_{i}(1),Y_{i}(0)\}$ mapped to treatment and control status, of which only the one corresponding to the observed treatment is observed, denoted by $Y_i=Z_{i}Y_{i}(1)+(1-Z_{i})Y_{i}(0)$; the other potential outcome is missing. The propensity score is the probability of a unit being assigned to the treatment group given the covariates \citep{rosenbaum1983central}: $e(x)=\Pr(Z_i=1\mid X_i=x)$.

The goal is to use the sample to estimate the average treatment effect (ATE): 
\begin{equation}
\tau^{\ATE} = \bE\{Y_{i}(1)-Y_{i}(0)\}=\bE_x\{\tau(x)\}, 
\end{equation}
where $\tau(x)=\bE\{Y_{i}(1)-Y_{i}(0)\mid X_i=x\}$ is the conditional average treatment effect at covariate value $x$. 
To identify $\tau^{\ATE}$, we {assume SUTVA \citep{rubin1980randomization} and} maintain the two standard assumptions: (i) unconfoundedness, $\Pr(Z_i\mid Y_i(0), Y_i(1), X_i)=\Pr(Z_i\mid X_i)$; (ii) overlap, $0<\Pr(Z_i=1\mid X_i)<1$. 
Among the consistent estimators of $\tau^{\ATE}$, we focus on the H\'{a}jek inverse probability weighting estimator for analytical simplicity:
\begin{equation}\label{eq:hajek}
    \widehat{\tau}_N = \frac{\sum_{i=1}^N Y_iZ_i/{e}(X_i)}{\sum_{i=1}^N Z_i / {e}(X_i)} - \frac{\sum_{i=1}^N Y_i(1 - Z_i)/\{1 - {e}(X_i)\}}{\sum_{i=1}^N (1 - Z_i) / \{1 - {e}(X_i)\}}.
\end{equation}
\cite{lunceford2004stratification} show that $N^{1/2}(\widehat{\tau}_N - \tau^{\ATE})$ is asymptotically normal with variance 
\begin{equation}\label{eq:hajek-var-truePS}
V = \mathbb{E}\left[\frac{[Y(1) - \mathbb{E}\{Y(1)\}]^2}{e(X)}\right] + \mathbb{E}\left[\frac{[Y(0) - \mathbb{E}\{Y(0)\}]^2}{1 - e(X)}\right].
\end{equation}
In observational studies, the true propensity score is unknown and is replaced by an estimated one. \cite{lunceford2004stratification} show that the asymptotic variance of the H\'{a}jek estimator with the propensity score estimated from {a parametric model}, $V_0$, is $V$ subtracted by a complex quadratic term and therefore smaller. Calculation of the quadratic term usually requires individual data, which is unavailable in power calculations. Therefore, we will proceed with $V$ henceforth, noting that it leads to a conservative (i.e. over-estimated) sample size estimate than that based on $V_0$. 
We investigate the discrepancy between $V$ and $V_0$ using simulations in Section \ref{sec:simulation}.  {Besides the inverse probability weighting estimators, other two classes of estimators of $\tau^{\ATE}$ are the outcome regression and doubly-robust estimators. However, asymptotic variances of these estimators depend on the specific outcome model, which is infeasible to pre-specify in the stage of power analysis. In contrast, as shown later, the H\'{a}jek estimator requires much weaker assumptions on the outcome model.}

With a fixed value of $V$,  a $(1 - \alpha)$ confidence interval for $\widehat{\tau}_N$ is
%\begin{equation}
$\widehat{\tau}_N \pm z_{1 - \alpha / 2} V^{1/2}N^{-1/2}$,
%\end{equation}
where $z_{1 - \alpha/2}$ is the $(1 - \alpha / 2)$ quantile of the standard normal distribution.  Without loss of generality, we consider a one-sided level-$\alpha$ test for the null hypothesis $H_0: \tau = 0$ against the alternative $H_a: \tau > 0$: 
\begin{equation}\label{eq:hypotest}
    \text{reject $H_0$ if $\widehat{\tau}_N > z_{1 - \alpha} ({V} / N)^{1/2}$.}
\end{equation} 
Standard derivation shows the minimal sample size $N$ to reach a desired power $\beta$ for a level-$\alpha$ test \eqref{eq:hypotest} with a fixed variance $V$ is (the proof is relegated to Appendix A.1):
    \begin{equation}\label{eq:sample-size-var}
        N = {\tau^{-2}}{\left(z_{1-\alpha} {V}^{1/2} - z_{1-\beta} {V_0}^{1/2}\right)^2} \leq {\tau^{-2}}{V}\left(z_{1-\alpha} + z_\beta\right)^2,
    \end{equation}
{assuming $\alpha < 0.5$ and $\beta > 0.5$, which commonly holds in practice.}

In a randomized experiment, $e(x) \equiv r$, where $r$ is the treatment proportion, so ${V}$ simplifies to {$\bV\{Y(1)\} / r + \bV\{Y(0)\} / (1 - r)$}. Moreover, $\Pr(Y(z))=\Pr(Y\mid Z=z)$, and thus the variance of the potential outcomes $Y(z)$ can be estimated by the sample variances of the outcome in arm $z$, $s_z^2$. Following the convention, we impose $s_0 = s_1=s$ and define the effect size as $\widetilde{\tau} = \tau / s$. Then, to detect an effect size $\widetilde{\tau}$ in a randomized experiment using a one-sided level-$\alpha$ test \eqref{eq:hypotest} with power $\beta$, the minimal sample size is 
{\begin{equation}\label{eq:ss-rct}
    N = \left(\frac{s_1^2}{r} + \frac{s_0^2}{1-r}\right)\frac{\left(z_{1-\alpha} + z_\beta\right)^2}{\tau^2} = \frac{\left(z_{1-\alpha} + z_\beta\right)^2}{r(1-r) \widetilde{\tau}^2}.
\end{equation}}
This coincides with the standard sample size formula of a two-sample $z$-test.  

In observational studies, power calculation is more complicated because (i) the propensity scores are not known, (ii) $\Pr(Y(z))\neq \Pr(Y\mid Z=z)$ due to confounding, and thus the terms involving potential outcomes, $\bE\{Y(z)\}$ and $\bV\{Y(z)\}$, deviate from their observed counterparts $\bE(Y\mid Z = z)$ and $\bV(Y\mid Z = z)$, and (iii) the numerator and denominator inside the expectation operator in \eqref{eq:hajek-var-truePS} can not be separated because the potential outcomes may depend on the propensity scores. When sampled units are available, we can readily address these complications. However, power calculation is conducted prior to data collection, with at best limited summary information to estimate the propensity scores or calculate the expectations.

\section{Sample size and power calculation} \label{sec:ss-calculation}
The essence of power calculation lies in computing the variance $V$ in \eqref{eq:hajek-var-truePS}. If we could observe $(X_i, Z_i, Y_i)$ for each sampled unit $i$, $V$ could easily be estimated by sample moments, but not so with only  summary data. To determine which summary information is sufficient for computing $V$, we rewrite the terms in expression \eqref{eq:hajek-var-truePS} as (taking $z=1$ for example), 
\begin{small}
\begin{eqnarray}
    \mathbb{E}\left[\frac{[Y(1) - \mathbb{E}\{Y(1)\}]^2}{e(X)}\right] = \mathbb{E}\left[{[Y(1) - \mathbb{E}\{Y(1)\}]^2}\right]\mathbb{E}\left\{\frac{1}{e(X)}\right\} 
    + \mathrm{Cov}\left[{[Y(1) - \mathbb{E}\{Y(1)\}]^2}, \frac{1}{e(X)}\right]. \nonumber
\end{eqnarray}
\end{small}
This expression suggests three components: the propensity score distribution, the potential outcome distribution, and their correlation. We study each component in detail below.

\subsection{Normal approximation of linear combinations of covariates}
Both the propensity score and the outcome models are commonly specified as a generalized linear model with a linear combination of covariates, $X'\beta = \sum_{j=1}^p \beta_j X_j$, as the predictor. However, in power calculations, we usually do not have specific information on each covariate. Previous authors often operated with a single normal predictor \cite[e.g.][]{raudenbush1997statistical, kelcey2017statistical}, but without theoretical justification. Here we invoke a special case of the Lyapunov Central Limit Theorem \citep{billingsley1995probability} to approximate the distribution of the scalar summary $X'\beta$ and thus justify such a procedure. 

\begin{theorem}\label{thm:normal_approx}
    Assume $X_1, X_2, \dots$ is a sequence of independent random variables with mean 0 and variance 1. Let $\beta_1, \beta_2, \dots$ be a sequence of real numbers. If the following two conditions hold: (i) there exists a positive number $B$ such that $\mathbb{E}(X_j^4) < B$, and (ii) $\max_{1 \leq j \leq p} {\beta_j^2}(\sum_{j=1}^p \beta_j^2)^{-1} = o(1)$, then 
    \begin{equation}
        \frac{1}{\sum_{j=1}^p \beta_j^2}\sum_{j=1}^p \beta_j X_j \Rightarrow \mathsf{N}(0, 1).
    \end{equation}
\end{theorem}
The proof is relegated to Appendix A.2.    Condition (i) ensures that the distribution of each covariate does not have heavy tails, and condition (ii) ensures that no single covariate dominates the linear combination. {Theorem \ref{thm:normal_approx} suggests that, under Condition (i)-(ii), the linear combination $X'\beta$ converges to a Normal distribution as the number of covariates ($p$) increases. In practice, some covariates might be correlated and $p$ might be modest. Some versions of the central limit theorem for dependent variables can be invoked when there is known correlation structure between covariates. However, in the context of power analysis, such information is usually not available and, in fact, analysts would avoid making specific assumptions about the covariate distributions. Instead, we conducted extensive simulations to examine the empirical distribution of $X'\beta$ under a wide range of scenarios. These simulations show $X'\beta$ approximates a Normal distribution satisfactorily often with as few as 10 variables as long as there is no strong pairwise correlation between $X$. We also draw the empirical distribution of $X'\beta$ in four real applications, including the benchmark RHC data used in Section \ref{sec:application}, with $p$ between 10 to 50, each of which also exhibits normality (details in Appendix B.1). Importantly, our derivation in the following sections only relies on the normality of $X'\beta$ rather than the exact distribution, which is easier to achieve with a modest $p$ and mild correlation between covariates.}   

\subsection{Distribution of propensity scores}\label{sec:dist-ps}
{This subsection shows how to, under a logistic model specification, uniquely identify the distribution of propensity scores with only two parameters: the treatment proportion $r$ and an overlap coefficient $\phi$ that measures the similarity of covariates between two groups. The key idea is to approximate the distribution of propensity scores by a computationally tractable Beta distribution, as elaborated below.}

Following the literature, we impose a logistic regression model for the propensity scores: $e(X) = \expit(W_e)$, where $W_e=\beta_0 + \sum_{j=1}^p \beta_j X_j$. Theorem \ref{thm:normal_approx} motivates us to approximate the distribution of the scalar summary $W_e$ by a Normal distribution $\mathsf{N}(\mu_e, \sigma_e^2)$, and thus approximate the propensity score distribution by a logit-normal distribution $\mathcal{P}(\mu_e, \sigma_e^2)$.   

To determine the two parameters $\mu_e$ and $\sigma_e$ of the logit-normal distribution, we need to impose constraints in the form of summary information about the treatment assignment. The first constraint, which is almost always available, is the proportion of the treatment group in the intended study: $r=\Pr(Z_i=1)$. The second constraint is a measure, denoted as $\phi$, of the similarity between the covariate distributions in the treatment and control groups, $\Pr(X\mid Z=1)$ versus $\Pr(X\mid Z=0)$. There are many candidates of $\phi$; important criteria include: (i) induce a 1-1 map between $(r,\phi)$ and $(\mu_e,\sigma_e)$ so that the generating process of the propensity score is unique, and (ii) easy to compute the propensity score distribution given $(r,\phi)$.   

The logit-normal distribution does not have a closed-form mean and thus it is generally difficult to solve for $(\mu_e,\sigma_e)$ given $(r, \phi)$. Instead, we propose to approximate the logit-normal distribution by a Beta distribution, which also has the support $[0,1]$ and is easy to compute.  Applying the method in \cite{aitchison1980logistic},  we can derive the ``optimal'' Beta distribution approximation of a logit-normal distribution by minimizing their Kullback-Leibler divergence. Specifically, for a logit-normal distribution $\mathcal{P}(\mu_e, \sigma_e^2)$ and its optimal Beta distribution approximation $\mathsf{Beta}(a, b)$, their parameters satisfy 
\begin{align}\label{eq:beta-logitnormal}
   \mu_e = \psi(a) - \psi(b), \quad  \sigma_e^2 = \psi'(a) + \psi'(b)
\end{align}
where $\psi$ and $\psi'$ are the digamma and trigamma functions, respectively. Numerical examples in Appendix B.2 show that this approximation hold for a wide range of $\mathcal{P}(\mu_e, \sigma_e^2)$.

If $e(X)\sim \mathsf{Beta}(a, b)$, then there is a closed-form relation between $a,b,r$: 
\begin{equation}\label{eq:r-ab}
    r = {a}/(a + b).
\end{equation} Therefore, it suffices to identify the propensity score distribution if we find another overlap measure $\phi$ easily computable from $a,b$. To this end, we look for a such $\phi$. Directly operating on the multivariate covariate $X$ is challenging; instead we operate on the propensity score $e(X)$, the coarsest balancing score \citep{rosenbaum1983central}. Let $f_z(u)$ be the density of $e(X)$ in arm $z$: $f_z(u) = f(e(X) = u \mid Z = z)$. We propose to use the Bhattacharyya coefficient \citep{bhattacharyya1943measure} to measure the covariate overlap between two arms: 
\begin{equation} \label{eq:overlap-def}
    \phi \equiv \int_0^1 \sqrt{f_1(u) f_0(u)} \,\mathrm{d} u.
\end{equation}
For two general densities, $0 \leq \phi \leq 1$ and $\phi = 0$ if and only if the support of $f_0$ and $f_1$ does not overlap, whereas $\phi = 1$ if and only if $f_0 = f_1$ almost everywhere. We term $\phi$ as the \emph{overlap coefficient} hereafter. In randomized trials, $f_0$ and $f_1$ are both point masses at $r$ and thus $\phi=1$, requiring no approximations. In observational studies where the Beta approximation is invoked, the Bhattacharyya coefficient has the major computational advantage in the following Proposition \ref{thm:phi-form}, satisfying our criteria for selecting the overlap measure.
% We choose the Bhattacharyya coefficient because it has a major computational advantage in observational studies, established in the following proposition. 
\begin{proposition}\label{thm:phi-form}
For $e(X)\sim \mathsf{Beta}(a, b)$ with $a,b>0$:
\begin{enumerate}
    \item [(i)] the overlap coefficient $\phi$ is a function of $(a,b)$ with the closed-form expression:
\begin{equation}\label{eq:phi-form}
   \phi =\frac{\Gamma(a + 0.5)} {{a}^{1/2}\Gamma(a)}\frac{\Gamma(b + 0.5)}{{b}^{1/2}\Gamma(b)};
\end{equation}
\vspace{-6pt}
\item [(ii)] given $r$, the $\phi$ is monotonically increasing in both $a$ and $b$;
\item [(iii)] there is a 1-1 map (bijection) between $(r,\phi) \in (0,1)^2$ and $(a,b)\in(0,\infty)^2$.
\end{enumerate}
\end{proposition}
The proof of (i)-(iii) is relegated to Appendix A.3-A.5, respectively. Proposition \ref{thm:phi-form} establishes that, given $(r, \phi)$, we can obtain the Beta parameters $(a, b)$ by solving two equations: (i) $r = {a}/(a + b)$, and (ii) Equation \eqref{eq:phi-form}. Furthermore, we can use the bisection method to solve for $(a,b)$ efficiently, based on which we reverse the Beta approximation of $e(X)$ to obtain the logit-normal parameters $(\mu_e,\sigma_e)$. 
It is worth noting that other specification of the overlap coefficient, e.g. the overlapping area between the density functions of two distributions \citep{inman1989overlapping}, usually does not lead to a closed-form expression of $\phi$ as a function of ($a,b$), which is crucial for verifying the 1-to-1 map between $(r, \phi)$ and $(a,b)$.

\begin{figure}
    \centering
    \includegraphics[width=0.95\linewidth]{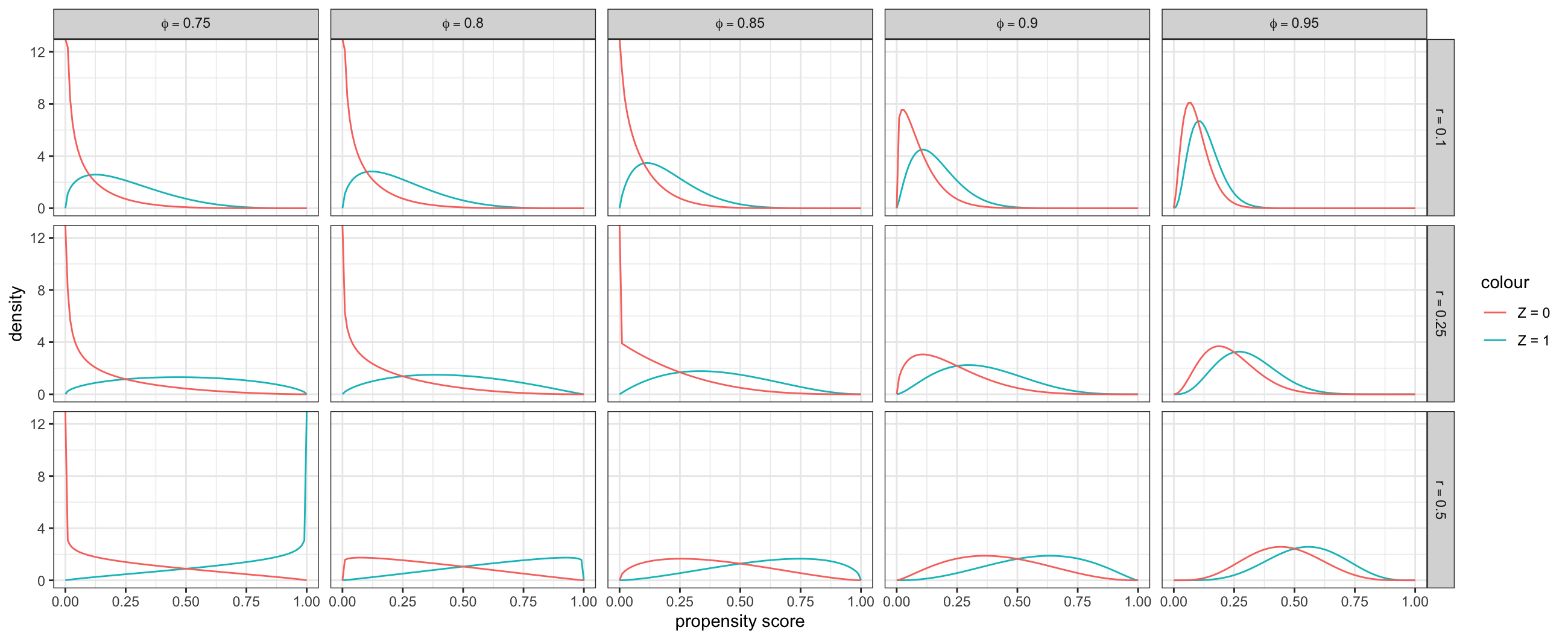}
    \caption{Distributions of $e(X)$ of the treatment and control groups corresponding to different combinations of $(r, \phi)$, where $e(X)$ follows a $\mathsf{Beta}(a, b)$ distribution with $a$ and $b$ determined by $(r, \phi)$.}
    \label{fig:overlap-coef}
\end{figure}

For $e(x) \sim \mathsf{Beta}(a, b)$ in observational studies, the $0 < \phi < 1$. We can use the Bayes theorem to show that $\Pr(e(x)\mid Z=1) \sim \mathsf{Beta}(a+1, b)$ and $\Pr(e(x)\mid Z=0) \sim \mathsf{Beta}(a, b+1)$. Therefore, as $\phi \to 1$, the propensity score distributions in the two groups converge to a common point mass at $r$, approaching a randomized trial. Figure \ref{fig:overlap-coef} shows the overlaying distributions of $e(X)$ in the $Z=1$ and $Z=0$ groups across combinations of $(r, \phi)$. Though $\phi$ has a theoretical range between 0 and 1, Figure \ref{fig:overlap-coef} shows that, however, there is little difference between the propensity score distribution if $\phi<0.8$, especially when $r<0.5$, the common treatment proportion in observational studies. Visual check reveals that, when $\min\{a,b\}<1$, the Beta distribution is U-shaped and the propensities are concentrated in the tail region (close to $0$ or $1$), suggesting extremely poor overlap between groups. Calculations shows  $a=b=1,1.5,2.5,5$ correspond to $\phi=0.80, 0.85, 0.90, 0.95$. Therefore, we set a simple heuristic rule of thumb for measuring the degree of overlap: $\phi<0.8, \in(0.8-0.9), \in (0.9-0.95), >0.95$ corresponds to very poor, poor, moderate, good overlap, respectively.  In practice, users may compute $\phi$ from a pilot study, of which we provide a computational method in Appendix B.3. When no pilot study is available, we recommend conducting sensitivity analysis on a range of $\phi$ based on domain knowledge.
 
\subsection{Distribution of potential outcomes} \label{sec:dist-po}

This subsection relates potential outcomes to propensity scores under mild model assumptions. The key idea is to project the outcomes into the space spanned by the scalar summary of the propensity score from Section \ref{sec:dist-ps}.

{
We assume a semiparametric partial linear or restricted mean model (RMM) with a homogeneous treatment effect $\tau$ for the potential outcomes 
\begin{equation}\label{eq:y-rmm}
    Y(z) = f(X) + z\tau + \epsilon.
\end{equation} 
Furthermore, we assume (i) homoscedasticity of the error, $ \bV\{\epsilon\mid e(X)\}=\bV(\epsilon)$, and (ii) $f(X)$ and $W_e$ are jointly Normal. We do not impose additional distributional assumptions on $Y(z)$ or $\epsilon$. The homoscedasticity assumption states that the propensity score is not predictive of the error variance of potential outcomes, and is weaker than the canonical version $\bV(\epsilon|X)=\bV(\epsilon)$; {invoking it avoids specifying the joint distribution of $Y(z)$ and $e(X)$ (for details, see derivations in Appendix A.7)}. The Normal assumption 
holds under various models for $f(X)$, most straightforwardly under a linear model $f(X) = \gamma_0 + \sum_{j=1}^p \gamma_j X_j$. Specifically, with a linear $f(X)$, we can approximate the joint distribution of $W_e$ and $f(X)$ by a bivariate normal distribution, invoking a multivariate extension of the Lyapunov central limit theorem \citep{bentkus2005lyapunov}. 
More generally, when $f(X)$ is itself an approximately normal function of covariates, as is often the case when no single covariate dominates as specified in Theorem \ref{thm:normal_approx}, the joint normality provides a reasonable approximation. We evaluate the empirical performance of our method under this assumption using simulations in Section~\ref{sec:simulation}.
}

{Under the joint normality of $f(X)$ and $W_e$, the correlation between the potential outcomes and the propensity score is captured by the correlation between $f(X)$ and $W_e$. Therefore, we decompose $f(X)$ as $f(X) = a W_e + W_e^\perp$, where $a W_e$ is collinear with $W_e$ and $W_e^\perp$ is orthogonal to $W_e$. Then, the potential outcomes can be expressed as $Y(z) = a W_e + W_e^\perp + z\tau + \epsilon$, where $W_e^\perp$ is independent of $W_e$ because $(W_e, W_e^\perp)$ is jointly Normal. By the definition of RMM, $\epsilon$ is also uncorrelated with any function of $X$, including $W_e$. Therefore, with a slight abuse of notation, we denote the compound error $W_e^\perp + \epsilon$ as $\epsilon$.
}

The centered potential outcome in $V$, i.e., $Y(z) - \mathbb{E}\{Y(z)\}$ for $z = 0, 1$, now reduces to $a\{W_e - \bE(W_e)\} + \{\epsilon - \bE(\epsilon)\}$, whose mean and variance are fully determined by two unknown parameters, $a$ and $\sigma_y^2=\bV(\epsilon)$. To solve $a$ and $\sigma_y^2$, two inputs suffice: $S^2=\bV\{Y(z)\}$, the variance of potential outcomes, and $\rho = \mathrm{cor}\{Y(z), W_e\}$, the correlation between potential outcomes and the linear predictor of propensity scores. Under the homogeneous effect model \eqref{eq:y-rmm}, the distribution of the centered potential outcome does not depend on $z$, nor do $\bV\{Y(z)\}$ or any correlation with $Y(z)$. One can obtain $(a, \sigma_y^2)$ from $(S^2, \rho)$ by the following equations (the proof is relegated to Appendix A.6):
\begin{align}
    a &= \rho |S|\sigma_e^{-1}, \quad \sigma_y^2 = (1 - \rho^2){S^2}.
    \label{eq:solve-unknowns}
\end{align}

The parameter $\rho$ measures the association between potential outcomes $Y(z)$ and $W_e$. Recall that the propensity score $e(X)=\expit(W_e)$; therefore, $W_e$ is a scalar summary of the covariates that are predictive of treatment assignment. Because confounders are covariates that are correlated with both treatment and outcome, $\rho$ summarizes the strength of confounding, which we term as the \emph{confounding coefficient}. It is interpretable in specific contexts; for example, it measures the strength of prognostic factors in medical studies and of self-selection in social sciences. The confounding coefficient $\rho$ is closely related to the R-squared statistic in the regression of $Y(z)$ on $X$, $R^2=\bV\{f(X)\} / \bV\{Y(z)\}$, which measures the proportion of outcome variance explainable by covariates. The following derivation shows that $\rho^2$ is bounded by $R^2$:
\begin{align}\label{eq:corr-parameter}
    \rho^2 
    &= \left[\frac{\mathrm{cov}\{Y(z), W_e\}}{\{\mathbb{V}(W_e)\}^{1/2} [\mathbb{V}\{f(X)\}]^{1/2}}\right]^2{\frac{\mathbb{V}\{f(X)\}}{\mathbb{V}\{Y(z)\}}} = \mathrm{cor}^2\{f(X), W_e\} R^2 \leq R^2.
\end{align}
In completely randomized trials, covariates are independent of the treatment assignment by design, so $W_e$ reduces to a constant, and thus $\mathrm{cor}\{f(X), W_e\}=0$ and $\rho=0$ by convention. In observational studies, $\mathrm{cor}\{f(X), W_e\}$ is not guaranteed to be zero, so $\rho^2$ is not either; instead, $\rho^2$ is bounded by $R^2$, a quantify also used in \cite{li2018asymptotic} to quantify covariate-outcome association. However, the bound is not tight; for example, in all synthetic and real-data-calibrated simulations in Section \ref{sec:simulation}, $\rho$ is close to 0 ($\rho<0.02$) whereas $R^2$ is much larger, often above $0.2$. This pattern is consistent with Equation \eqref{eq:corr-parameter}, because $\mathrm{cor}\{f(X), W_e\}$ is usually small unless the propensity score itself is a strong predictor of potential outcomes. In practice, one may estimate $\rho^2$ from a pilot study, or conduct a sensitivity analysis over a small range, e.g. $\rho^2 \in (0, 0.05)$, or simply set $\rho=0$, as implicitly imposed in \cite{shook2022power}, which may mildly underestimate the sample size.

\subsection{Analytical formula for sample size calculation}\label{sec:ss-formula} 
{In this subsection, we build on Sections \ref{sec:dist-ps}-\ref{sec:dist-po} to derive an analytical sample size calculation formula. }
Following the convention of power analysis, we define the effect size as the treatment effect standardized by the standard error: $\widetilde{\tau} = \tau / S$. The following theorem presents the analytical formula for the sample size calculation of $\tau^{\ATE}$ (the proof is relegated to Appendix A.7).

\begin{theorem}\label{thm:sample-size-formula}
Let $\widetilde{\tau}$ be the desired standardized effect size of the ATE in an observational study with a treatment proportion of $r$, an overlap coefficient $\phi \in (0,1]$, and a confounding coefficient $\rho\in[0,1)$. The minimum sample size for the one-sided level-$\alpha$ test defined in \eqref{eq:hypotest} for detecting $\widetilde{\tau}$ with power $\beta$ is $N = \widetilde{V}\left(z_{1-\alpha} + z_\beta\right)^2/{\widetilde{\tau}^2}$, with
    {
    \begin{equation} 
        \widetilde{V} = 2\left\{1 + (\rho^2\sigma_e^2 + 1)\exp\left(\frac{\sigma_e^2}{2}\right)\cosh(\mu_e)\right\},\label{eq:ss-formula}
    \end{equation}
    where $\cosh(\mu_e)=\{\exp(\mu_e)+\exp(-\mu_e)\}/2$, and $(\mu_e, \sigma_e^2)$ are uniquely determined by $(r, \phi)$ through Equation \eqref{eq:beta-logitnormal}, \eqref{eq:r-ab}, \eqref{eq:phi-form}. 
    }
   \end{theorem}
The minimum sample size for the two-sided counterpart of test \eqref{eq:hypotest} is obtained by replacing $z_{1-\alpha}$ with $z_{1-\alpha/2}$ in Theorem~\ref{thm:sample-size-formula}. Equation \eqref{eq:ss-formula} shows that the variance $\widetilde{V}$ depends only on $\rho$ and $(\mu_e, \sigma_e^2)$, where the latter is solved from the distribution of $e(X)$, which is uniquely determined by the inputs $(r, \phi)$. In a randomized trial, the propensity score is fixed at $e(x)=r$ with $\phi=1$ and $\rho=0$, in which $\mu_e$ converges to $\log\{r/(1-r)\}$ and $\sigma_e$ converges to $0$ so that formula \eqref{eq:ss-formula} reduces to the $z$-test-based formula \eqref{eq:ss-rct}. Theorem~\ref{thm:sample-size-formula} reduces the number of inputs for the sample size calculation to two, namely $(\phi, \rho)$, in addition to the standard $\widetilde{\tau}, \alpha, \beta, r$. 
Based on Equation \eqref{eq:ss-formula} and jointly with Proposition \ref{thm:phi-form}(ii) and the relation in \eqref{eq:beta-logitnormal}, the following corollary on the relationship between $N$ and $(\rho, \phi)$ is immediate.   

\begin{corollary} \label{thm:ss-monotone}  The variance $\widetilde{V}$ and thus the sample size $N$ (i) monotonically increases in $|\rho| \in (0,1]$ given  $(r, \phi)$, and (ii) monotonically decreases in $\phi\in(0,1]$ given $r$.
\end{corollary}

Figure \ref{fig:flowchart} shows a flowchart of how to decompose the variance of the H\'{a}jek estimator into three components and compute each component from the proposed inputs.

\tikzset{
  block/.style={
    rectangle, draw, text width=#1, text centered,
    inner sep=4pt, font=\footnotesize, align=center
  },
  block/.default=7.5em,
  arr/.style={-{Stealth[length=5pt]}, thick},
}
\begin{figure}[ht]
\begin{tikzpicture}[scale=0.9]

% === Row 1 ===
\node[block=10em] (comp) at (0, 0) {Components needed for computation};
\node[block=20em] (dist) at (8.5, 0) {Distribution of propensity score, potential outcome, and correlation between them};

% === Row 2 ===
\node[block=10em] (ps) at (4, -2.8) {Propensity score:\\logit-normal $\longleftrightarrow$ Beta};
\node[block=10em] (po) at (8.5, -2.8) {Potential outcome:\\normal distribution};
\node[block=9em] (corr) at (13, -2.8) {Correlation:\\$\rho = \mathrm{cor}\{W_e, Y(z)\}$};

% === Row 3 ===
\node[block=7.5em] (how) at (0, -5.5) {How to determine the parameters?};
\node[block=8em] (tp1) at (4, -5.5) {Two parameters.\\Two inputs:\\i.\ trt proportion $r$\\ii. {overlap $\phi$}};
\node[block=10em] (tp2) at (8.5, -5.5) {Two parameters.\\Two inputs:\\i.\ $S^2 = \mathrm{Var}\{Y(z)\}$\\ii.\ $\rho = \mathrm{cor}\{W_e, Y(z)\}$};
\node[block=9em] (cv) at (13, -5.5) {Sensitivity parameter.\\$\rho^2 \leq R^2$\\(bounded by $R$-squared)};

% === Arrows ===
\draw[arr] (comp.east) -- (dist.west);
\draw[arr] (comp.south) -- (how.north);
\draw[arr] (dist.south) -- (ps.north);
\draw[arr] (dist.south) -- (po.north);
\draw[arr] (dist.south) -- (corr.north);
\draw[arr] (how.east) -- (tp1.west);
\draw[arr] (ps.south) -- (tp1.north);
\draw[arr] (po.south) -- (tp2.north);
\draw[arr] (corr.south) -- (cv.north);
\draw[arr] (tp1.east) -- (tp2.west);
\draw[arr] (tp2.east) -- (cv.west);
\end{tikzpicture}
\caption{Flow chat of decomposing and computing the variance of the H\'{a}jek estimator from summary inputs.}
\label{fig:flowchart}
\end{figure}

{\textbf{Remark 1.} In summary, our derivation relies on (i) a logistic model for the propensity scores, and (ii) a semiparametric homoscedastic restricted mean outcome model with a homogeneous treatment effect. In real applications, the true data generating process might deviate from the assumptions. In particular, heterogeneous treatment effects are common. We can extend the homogeneous outcome model \eqref{eq:y-rmm} to a varying coefficient model to accommodate heterogeneity: $Y(z) = f(X) + z\tau(X) + \epsilon$, where $\tau(X)$ stands for the conditional average treatment effect at covariate $X$. Maintain the homoscedasticity and joint normality assumptions and use the same projection to the space spanned by $W_e$ as in the homogeneity model. Then, simple algebra (details are omitted) shows that the asymptotic variance of the H\'ajek estimator $V$ of the heterogeneity model depends on the variance of $\tau(X)$, the knowledge of which is generally not available before data collection as in power analysis. This challenge is also discussed in the context of sample size calculation for rerandomization \citep{branson2024power}. Instead, we use simulations in Section \ref{sec:simulation} to examine consequences of violation to homogeneity and find Formula \eqref{eq:ss-formula} is generally robust. }    

\textbf{Remark 2.} Our analytical derivation elucidates the connection and distinction between power calculations in randomized trials and observational studies. Specifically, both cases require four parameters: $\alpha$, $\beta$, the standardized effect size $\widetilde{\tau}$, and the treatment proportion $r$. Observational studies require two additional inputs: the overlap coefficient $\phi$ and the correlation parameter $\rho^2$, both of which have a natural interpretation. This result is intuitive because the key difference between observational studies and randomized trials is the presence of confounders in the former, which are pre-treatment covariates that affect the treatment assignment and the outcome simultaneously. The parameters $\phi$ and $\rho^2$ quantify the strength of confounder-treatment and confounder-outcome association, respectively.
Intuitively, as the covariate overlap between groups (encoded by $\phi$) decreases and the correlation between covariates and the outcome (encoded by $\rho^2$) increases, an observational study deviates further from a randomized trial, rendering the marginal distribution of the observed outcomes deviate further from that of the potential outcomes and thus require more sample units to identify the causal estimands. {This matches Corollary \ref{thm:ss-monotone}.}

\textbf{Remark 3.} The sample size formula in  \cite{shook2022power}, derived based on a variance inflation factor, is a special case of formula \eqref{eq:ss-formula}. Specifically, assuming $e(X) \sim \expit(\mathsf{N}(\mu_e, \sigma_e^2))$, {their formula becomes $\widetilde{V}_{\mathrm{SH}} = 2\left\{1 + \exp(\sigma_e^2 / 2)\cosh(\mu_e)\right\}$. Comparing to formula \eqref{eq:ss-formula}, we have $\widetilde{V}=\widetilde{V}_{\mathrm{SH}}+2\rho^2\sigma_e^2\exp(\sigma_e^2 / 2)\cosh(\mu_e)$. Therefore, 
$\widetilde{V}_{\mathrm{SH}}$ equals $\widetilde{V}$ when $\rho=0$ and is smaller than $\widetilde{V}$ otherwise.} Although $\rho^2$ is small in many of our simulation studies, ignoring it risks underestimating the sample size in general settings. \cite{shook2022power} discussed a term that implicitly adjusts for the confounder-treatment association, but it is not estimable without prior data on both weights and outcomes.  

\textbf{Remark 4.} For binary outcomes, because the mean of a Bernoulli distribution determines the variance, compared to  continuous outcomes in Section \ref{sec:dist-po}, we have two fewer constraints to determine the outcome distribution. As a simple solution, we recommend using linear models for binary outcomes so that the proposed method directly applies. Numerical examples in Section \ref{sec:application} and Appendix C.1 suggest that this approach generally works well in practice.

\section{Weighted average treatment effect estimands} \label{sec:WATE}
In the presence of poor overlap between the comparison groups, the inverse probability weight can lead to excessive variance due to extreme propensity scores. Alternative weighting schemes are often more desirable. In this section, we extend the above methodology to the general class of weighted average treatment effect (WATE). Assume that the observed sample is drawn from a population with covariates distribution $f(x)$, and let $f(x)h(x)$ denote the covariate distribution of a \emph{target population}, where $h(x)$ is called a tilting function. Then, we can represent the average treatment effect on that target population using a WATE:  $\tau^h={\bE\{\tau(x)h(x)\}}/{\bE\{h(x)\}}$. When $h(x)=1$, $\tau^h$ reduces to ATE; when $h(x)=e(x)$, $\tau^h$ is the average treatment effect for the treated (ATT); when $h(x)=e(x)(1-e(x))$, $\tau^h$ becomes the average treatment effect for the overlap population (ATO).
We can use a unified weighting strategy to identify the WATE because $\tau^h=\bE\{w_1(X)ZY-w_0(X)(1-Z)Y\}$, where $w_1(x)=h(x)/e(x), w_0(x)=h(x)/(1-e(x))$ \citep{li2018balancing}. This corresponds to (i) the inverse probability weight (IPW), $(w_1=1/e(x), w_0=1/(1-e(x)))$, for the ATE; (ii) the ATT weight, $(w_1=1, w_0=e(x)/(1-e(x)))$, for the ATT; and (iii) the overlap weight (OW), $(w_1=1-e(x), w_0=e(x))$, for the ATO. 

The H\'{a}jek estimator for the WATE is:
\begin{equation}
\label{eq:Hajek-WATE}
\widehat{\tau}_w=\frac{\sum_i w_1(X_i)Z_i Y_i}{\sum_i w_1(X_i)Z_i} -
              \frac{\sum_i w_0(X_i)(1-Z_i) Y_i}{\sum_i w_0(X_i)(1-Z_i)}.
\end{equation}
We follow Section \ref{sec:ss-calculation} to derive sample size formulas for the ATT and ATO estimands. Specifically, the procedures to determine the propensity score and the outcome distributions remain the same, but the variance formula changes. Unlike the ATE, there is no deterministic order between the variance of the H\'{a}jek estimator with the estimated versus true propensity score for other WATE estimands \citep{yang2018asymptotic}. For the same reasons as those for the ATE discussed in Section \ref{sec:variance}, we proceed with the variance with the true propensity score.  

We show the variance of the H\'{a}jek estimator for the WATE with the tilting function $h$ is:
\begin{equation} \label{eq:WATE-var}
    V_w \equiv \bV(\widehat{\tau}_w)=\frac{1}{\mathbb{E}\{h(X)\}^2}\mathbb{E}\left[\left\{\frac{(Y(1) - \xi_1)^2}{e(X)} + \frac{(Y(0) - \xi_0)^2}{1 - e(X)} \right\} h(X)^2\right],
\end{equation}
where $\xi_z = \mathbb{E}_h\{Y(z)\} = \mathbb{E}\{h(X)Y(z)\} / \mathbb{E}\{h(X)\}$; the proof is in Appendix A.8. For the ATE, $h(X) = 1$, and $V_w$ reduces to $V$ in equation \eqref{eq:hajek-var-truePS}. For the ATT and ATO, $h(X)$ is a function of $e(X)$, and thus $V_w$ is also determined by the joint distribution of $e(X)$ and $Y(z)$, which can be computed as in Sections \ref{sec:dist-ps} and \ref{sec:dist-po}. However, neither the ATT nor the ATO has a closed-form expression of variance, but they can be computed numerically.

Specifically, for ATT and ATO, because $e(X)=\expit(W_e)$ under the logistic propensity model, so we can express $h(X)$ by $h(W_e)$, where $h(W_e) = \expit(W_e)$ for ATT and $h(W_e) = \expit(W_e)[1 - \expit(W_e)]$ for ATO. We further show that $V_w=S^2 \cdot \widetilde{V}_w$, where
\begin{align}\label{eq:sample-size-WATE}
  \widetilde{V}_w = \frac{1}{\bE[h(W_e)]^2} \left[ \frac{\rho^2}{\sigma_e^2} \{\mathcal{A}_1(h)+ \mathcal{A}_0(h)\} + (1-\rho^2) \{\mathcal{B}_1(h)+ \mathcal{B}_0(h)\} \right]
\end{align}
is the variance of the H\'{a}jek estimator of the standardized effect $\widetilde{\tau}=\tau^h/S$, and $\mathcal{A}_z(h)= \bE[(W_e-\zeta)^2 g_z(W_e)]$, $\mathcal{B}_z(h)= \bE[g_z(W_e)]$, $g_z(W_e)={h(W_e)^2}[z\expit(W_e)+(1-z)\{1-\expit(W_e)\}]^{-1}$, and  $\zeta = \bE[h(W_e)W_e]/\bE[h(W_e)]$; the proof is in Appendix A.9. All terms only involve $W_e$, therefore, we can approximate $\widetilde{V}_w$ of ATT and ATO using the same design-stage input $(r, \phi, \rho)$, and the sample size for a one-sided level-$\alpha$ test at power $\beta$ to detect the effect $\widetilde{\tau}$ is $N=\widetilde{V}_w \left(z_{1-\alpha} + z_\beta\right)^2/{\widetilde{\tau}^2}$. Evaluating these expectations requires numerical integrations that are standard in statistical software.

We summarize these steps in the following Algorithm \ref{alg:approx_Vw}.
\begin{algorithm}[Approximation of $\widetilde{V}_w$]\label{alg:approx_Vw}
\textit{Input}: $r$, $\phi$, $\rho$.
\begin{enumerate}[itemindent=0em]
    \item Solve the Beta distribution of $e$ with $(r,\phi)$ by Proposition \ref{thm:phi-form}.
    \item Evaluate $\mathbb{E}[h(W_e)]^2$, $\mathcal{A}_z(h)$, and $\mathcal{B}_z(h)$ for $z=0,1$ numerically by using $W_e = \logit{(e)}$, where the distribution of $e$ is solved in Step 1. Calculate $\sigma_e^2=\bV[W_e]$.
    \item Calculate $\widetilde{V}_w$ by plugging in quantities obtained from Step 2 and $\rho$ into equation \eqref{eq:sample-size-WATE}.
\end{enumerate}
\end{algorithm}

\section{Numerical Experiments} \label{sec:simulation}
\subsection{Synthetic Data: homogeneous treatment effect} \label{sec:simulation-design}
Following the simulation design in \cite{li2019addressing}, we simulate $N = 1,000,000$ units with 10 covariates as the superpopulation. Covariates $X_1$ to $X_4$ are binary with $\mathsf{Ber}(p)$ with $p = 0.2, 0.4, 0.6, 0.8$, respectively; $X_5 \sim \mathsf{Unif}(0, 1)$; $X_6$ to $X_8$ follows a Poisson distribution with mean parameters $1, 2,$ and $3$ respectively; $X_9 \sim \mathsf{Gamma}(2, 3)$ and $X_{10} \sim \mathsf{Beta}(2, 3)$. We generate the treatment indicator $Z \sim \mathsf{Ber}(\expit(\beta_0 + \kappa X \beta))$, where $\beta = (1, 1, -1, 0, -2, 1, 0.5, 0, 0, 0)'$. The parameter $\kappa \in \{0, 0.25, 0.5, 0.75, 0.9, 1\}$ controls the overlap between two groups, corresponding to the overlap coefficient $\phi \in \{1.00, 0.98, 0.93, 0.87, 0.84, 0.81\}$, respectively; $\phi = 1$ yields a randomized trial. Appendix C.1 visualizes the overlap between two groups. Correspondingly, we set $\beta_0 \in \{0, -0.248, -0.489, -0.722, -0.860, -0.951\}$ so that the treatment proportion $r = \Pr(Z=1) = 0.5$. We simulate from a potential outcome model with a \emph{homogeneous treatment effect} $\tau$: $Y(z) \sim \mathsf{N}(X\gamma + \tau z, 4^2)$, where $\gamma = (1, 1, -1, -1, 0, -1, -1, 0, 1, 1)'$ and $\tau=1$, and let the observed outcome $Y_i=Z_iY(1)+(1-Z_i)Y(0)$. We also simulate binary outcomes and relegate the results to Appendix C.2.

We fix the nominal power at $0.80$ and solve the sample size for each $\kappa$. We calculate the summary statistics $(r, \phi, S,\rho)$ as defined in Section \ref{sec:ss-calculation} from the superpopulation. The treatment proportion $r$ is fixed at $0.5$ throughout. We fit a logistic propensity score model, $e(X) = \expit(\beta_0 + X\beta)$, and the latent variable $W_e$ is estimated by $\widehat{\beta}_0 + X \widehat{\beta}$, where $(\widehat{\beta}_0, \widehat{\beta})$ are the estimated intercept and coefficients. Although none of the covariates is normal, $W_e$, the latent linear combination of covariates, appears to be close to normal (see the rightmost column of Figure 1 in Appendix B.1). We calculate the pooled variance $S^2$ and correlation $\rho$ empirically. To  calculate $\phi$, we use the empirical cumulative probability function of the propensity scores (details are relegated to Appendix B.3). We present $\phi, \rho^2, R^2$ in Table \ref{tab:sim1-result}.  The R-squared statistic $R^2$ is obtained by fitting the linear model $Y \sim X$ to the simulated data, with $Y$ centered within the respective treatment arm. The $R^2$ turns out to be much larger than $\rho^2$, which is not surprising as $R^2$ is the theoretical upper bound of $\rho^2$, and provides the most conservative estimate when there is no substantive knowledge on  $\rho^2$.

\begin{table}[htbp]
    \centering
    \caption{Sample size calculation by the proposed method, the two-sample $z$-test and the method in \cite{shook2022power} to achieve nominal power of 0.80 under various degrees of overlap. The power is obtained from random sampling of the calculated size from the $N = 1,000,000$ units. The Monte Carlo error for a power of 0.8 is 0.004. The treatment proportion $r$ is fixed at 0.5. The empirical $S^2$ is close to $20$ in all settings.}
    \label{tab:sim1-result}
    \begin{tabular}{cccccccccccc}
    \toprule
      & &&\multicolumn{3}{c}{Proposed} & \multicolumn{3}{c}{$z$-test} & \multicolumn{3}{c}{Shook-sa \& Hudgens} \\
      \cmidrule(lr){4-6} \cmidrule(lr){7-9} \cmidrule(lr){10-12}
      $\phi$ & $\rho^2$& $R^2$& Size & True PS & Est PS & Size & True PS & Est PS & Size & True PS & Est PS \\ \midrule
      1.00 &0.02 &0.20& 645 & 0.83 & 0.88 & 629 & 0.80 & 0.87 & 629 & 0.80 & 0.87 \\
      0.98 &0.03 &0.20& 656 & 0.79 & 0.87 & 628 & 0.78 & 0.85 & 653 & 0.80 & 0.88 \\
      0.93 &0.03 &0.19& 754 & 0.79 & 0.87 & 626 & 0.73 & 0.80 & 742 & 0.78 & 0.86 \\
      0.88 &0.02 &0.19& 993 & 0.78 & 0.85 & 624 & 0.60 & 0.65 & 950 & 0.77 & 0.84 \\
      0.84 &0.02 &0.19& 1290 & 0.80 & 0.85 & 623 & 0.48 & 0.53 & 1216 & 0.77 & 0.84 \\
      0.81 &0.02 &0.19& 1617 & 0.81 & 0.87 & 622 & 0.40 & 0.45 & 1481 & 0.79 & 0.81 \\ 
      \bottomrule
    \end{tabular}
\end{table}

We use inputs $(r, \phi, \rho^2)$ and the standardized effect $\widetilde{\tau} = 1 / S$ to inversely calculate the required sample size $N_{\mathrm{pos}}$ to achieve the nominal power $0.80$. Then, we sample $N_{\mathrm{pos}}$ units randomly from the $N$ units in the superpopulation. We repeat the process for $B = 10,000$ times; each time, we conduct the hypothesis test \eqref{eq:hypotest} at level $\alpha = 0.05$, with $\widehat{\tau}_N$ calculated with true and estimated propensity scores, respectively. We then estimate the power by the proportion of rejecting the null among the $B$ replicates. Results are reported in Table \ref{tab:sim1-result}, which also includes the method of \cite{shook2022power} (SH). The proposed method consistently attains powers close to the nominal 0.80; the SH method leads to slightly lower power. The similarity between the results of our proposed and the SH methods is as expected from Remark 3 given the small $\rho^2$.  As noted before, the power based on the estimated propensity scores is higher than that based on the true scores, rendering the proposed method conservative. {The discrepancy in the variance between the true and estimated propensity scores could be magnified in simulations compared to most real studies, because both $Z$ and $Y$ are drawn from the underlying true model.} We also include the sample size from a two-sample $z$-test for comparison. When overlap is good, the $z$-test correctly calculates the sample size. However, when the overlap is poor, the $z$-test, which treats observational studies as randomized trials, tends to vastly underestimate the sample size. 

\begin{table}[ht]
    \centering
    \caption{Sample size (size) calculated by the proposed method to achieve the nominal power 0.8 under various degrees of overlap. The power is obtained from bootstrapping 10,000 random samples of the calculated size.}
    \label{tab:sim1-result-ATT-ATO}
    \begin{tabular}{cccccccc}
    \toprule
      & \multicolumn{1}{c}{ATT} & \multicolumn{2}{c}{Power} & &\multicolumn{1}{c}{ATO} & \multicolumn{2}{c}{Power} \\ \cmidrule(lr){2-2} \cmidrule(lr){3-4} \cmidrule(lr){6-6} \cmidrule(lr){7-8}
      $\phi$ & Size & True PS & Est PS & & Size & True PS & Est PS \\ \midrule
      1.00 & 630 & 0.80 & 0.87 & & 630 & 0.80 & 0.88 \\
      0.98 & 680 & 0.79 & 0.86 & & 653 & 0.80 & 0.88 \\
      0.93 & 860 & 0.80 & 0.84 & & 711 & 0.80 & 0.88 \\
      0.88 & 1292 & 0.82 & 0.82 & & 796 & 0.82 & 0.88 \\
      0.84 & 1856 & 0.86 & 0.84 & & 877 & 0.84 & 0.88 \\
      0.81 & 2407 & 0.87 & 0.85 & & 918 & 0.85 & 0.88 \\ 
      \bottomrule
    \end{tabular}
\end{table}

Table \ref{tab:sim1-result-ATT-ATO} reports the required sample sizes for the ATT and ATO estimands under the same simulation design. The empirical power closely tracks the nominal across all degrees of overlap, paralleling the ATE results. The three estimands, however, diverge sharply as overlap deteriorates. The sample sizes for ATE and ATT both grow steeply as $\phi$ decreases. In contrast, the sample size for ATO grows only modestly. This provides a clear implication for study design: when investigators anticipate poor covariate overlap and the scientific question can be framed on a population at clinical equipoise, ATO offers a substantially more efficient estimand, and overlap weighting should be preferred over inverse probability weighting.

The confounding coefficient $\rho^2$ is usually unknown; we suggest treating it as a sensitivity parameter in a range close to 0. For $\phi \in \{1.00, 0.98, 0.93, 0.88, 0.84, 0.81\}$, sensitivity analyses yield sample size intervals $(629,630)$, $(653,669)$, $(742,821)$, $(959,1239)$, $(1225,1818)$ and $(1515,2516)$, respectively. Under good overlap, the required sample size is robust to the choice of $\rho^2$; as overlap deteriorates, sensitivity to $\rho^2$ grows sharply, resulting in wider intervals. This highlights the role of $\rho^2$, which has been ignored in the existing literature. More importantly, it reinforces the critical role of covariate overlap in observational study design: poor overlap not only inflates the required sample size but also amplifies its sensitivity to $\rho^2$, the strength of confounding.

\subsection{{Synthetic Data: heterogeneous treatment effect}} \label{sec:simu-hte}

{We extend the simulation in Section~\ref{sec:simulation-design} to assess the proposed method in the presence of heterogeneous treatment effect (HTE). We generate covariates, propensity score, treatment indicator, and the control potential outcome $Y_i(0)$ following Section~\ref{sec:simulation-design}, and simulate $N=1,000,000$ units as the superpopulation. Following the treatment effect decomposition in \cite{DingFeller2019HTE}, we generate the individual treatment effect as $\tau_i = \tau_0 + g(X_i) + u_i$, where $g(X_i)$ is the systematic component driven by observed covariates and $u_i \sim \mathsf{N}(0, 0.2^2)$ is the idiosyncratic component independent of observed covariates. The treated potential outcome is $Y_i(1) = Y_i(0) + \tau_i$. We target the ATE estimand $\tau=\bE[\tau_i]$. We consider two systematic components: (i) covariate-driven HTE, where $g(X_i) = \delta(X_{1,i} + X_{9,i})$, so that treatment effect heterogeneity is driven by both a confounder and a precision variable, and (ii) propensity score-driven HTE, where $g(X_i)=\delta W_{e,i}$, so that treatment effect varies along the logit propensity score. For each scenario, we calibrate $(\tau_0, \delta)$ so that $\bE[\tau_i]=\bV(\tau_i) = 1$ in the superpopulation, representing substantial effect heterogeneity where the standard deviation of individual treatment effects equals the true ATE. $\phi=1$ ($\kappa=0$) is not applicable to propensity score-driven HTE because $W_e$ is constant, and is thus omitted. By construction, this HTE mechanism increases $\bV[Y(1)]$ while $\bV[Y(0)]$ remains unchanged, a feature also shared by control potential outcome-driven HTE simulation designs in \cite{DingFeller2015HTE, Zach2020samplingRand}.} 

{As in Section~\ref{sec:simulation-design}, from the superpopulation, we estimate $W_e$ by $\widehat{W}_e$ with a logistic model and $\tau$ by $\hat{\tau}^{\OW}$ with the overlap weight estimator. Under the homogeneity assumption, we impute all $Y_i(0)$ for the group $Z_i=1$ by $Y_i-\hat{\tau}^{\OW}$ and all $Y_i(1)$ for the group $Z_i=0$ by $Y_i+\hat{\tau}^{\OW}$. Then, we compute the pooled variance $S^2=\bV[Y(0)]$, correlation $\rho=\mathrm{cor}(Y(0),W_e)$, and R-squared statistic $R^2=\bV[f(X)]/\bV[Y(0)]$ empirically. Although $\tau_i$ varies individually, this homogeneous imputation ensures empirical $S^2$, $\rho$, and $R^2$ calculated with imputed $Y(z)$ are the same for $z=0,1$ in the superpopulation. The $(r, \phi)$ remains unchanged from Section~\ref{sec:simulation-design}. Similarly, we use $(r,\phi,\rho^2)$ and the standardized effect $\tilde{\tau}=1/S$ to calculate the required sample size $N_{\mathrm{pos}}$ and estimate the empirical power through $B = 10,000$ replications. The empirical power is the proportion of rejecting the null hypothesis in test~\eqref{eq:hypotest} at level $\alpha=0.05$ with the H\'{a}jek estimator over $B$ replications. We report the simulation results in Table~\ref{tab:hte-result-ATE}. We also conduct simulations using the same simulated superpopulations under ATT and ATO estimands, and results are relegated to Appendix C.3.}

\begin{table}
    \centering
    \caption{Sample size calculation by the proposed method, the two-sample $z$-test and the method in \cite{shook2022power} to achieve nominal power of 0.80 under various degrees of overlap and HTE. The power is obtained from random sampling of the calculated size from the $N = 1{,}000{,}000$ units. The Monte Carlo error for a power of 0.8 is 0.004. The empirical $S^2$ is about 21.1 and 19.5 for the upper and lower panels, respectively.}
    \label{tab:hte-result-ATE}
    \begin{tabular}{cccccccccccc}
    \toprule
      &&& \multicolumn{3}{c}{Proposed} & \multicolumn{3}{c}{$z$-test} & \multicolumn{3}{c}{Shook-Sa \& Hudgens} \\
      \cmidrule(lr){4-6} \cmidrule(lr){7-9} \cmidrule(lr){10-12}
      $\phi$ & $\rho^2$& $R^2$& Size & True PS & Est PS & Size & True PS & Est PS & Size & True PS & Est PS \\ \midrule
      \multicolumn{12}{c}{\textit{covariates-driven HTE}} \\
      1.00 & 0.02 & 0.23 & 663 & 0.82 & 0.89 & 663 & 0.82 & 0.89 & 663 & 0.81 & 0.88 \\
      0.98 & 0.03 & 0.23 & 693 & 0.80 & 0.88 & 664 & 0.80 & 0.88 & 691 & 0.82 & 0.88 \\
      0.93 & 0.03 & 0.24 & 795 & 0.81 & 0.88 & 665 & 0.72 & 0.81 & 787 & 0.81 & 0.88 \\
      0.87 & 0.03 & 0.24 & 1048 & 0.81 & 0.87 & 666 & 0.61 & 0.67 & 1010 & 0.78 & 0.84 \\
      0.83 & 0.03 & 0.24 & 1363 & 0.80 & 0.85 & 666 & 0.49 & 0.53 & 1253 & 0.75 & 0.81 \\
      0.81 & 0.03 & 0.24 & 1715 & 0.79 & 0.84 & 667 & 0.37 & 0.42 & 1528 & 0.75 & 0.74 \\ \midrule
      \multicolumn{12}{c}{\textit{propensity score-driven HTE}} \\
      0.98 & 0.01 & 0.17 & 641 & 0.79 & 0.86 & 616 & 0.79 & 0.85 & 640 & 0.80 & 0.86 \\
      0.93 & 0.00 & 0.17 & 727 & 0.80 & 0.85 & 613 & 0.72 & 0.78 & 726 & 0.79 & 0.85 \\
      0.87 & 0.00 & 0.17 & 942 & 0.78 & 0.82 & 612 & 0.60 & 0.63 & 928 & 0.77 & 0.83 \\
      0.83 & 0.00 & 0.17 & 1207 & 0.78 & 0.82 & 611 & 0.49 & 0.47 & 1150 & 0.77 & 0.80 \\
      0.81 & 0.00 & 0.17 & 1499 & 0.73 & 0.78 & 611 & 0.39 & 0.37 & 1401 & 0.71 & 0.76 \\
      \bottomrule
    \end{tabular}
\end{table}

{Table~\ref{tab:hte-result-ATE} reports the calculated sample sizes and empirical power under HTE. Remark~1 notes that under HTE, the asymptotic variance of the H\'ajek estimator additionally depends on $\bV[\tau(X)]$, which is generally unavailable at the design stage. Nevertheless, the proposed method continues to achieve empirical power close to the nominal 0.80 across almost all overlap levels and both HTE scenarios, even when $\bV(\tau_i)$ equals the true ATE. Similar to the case of homogeneous effect, the power with estimated propensity scores exceeds, often by a sizeable margin, that with the true scores under HTE. The comparison with the $z$-test and the method of \cite{shook2022power} mirrors the homogeneous case: the $z$-test increasingly underestimates the required sample size as overlap deteriorates, while the SH method tracks the proposed method closely when $\rho^2$ is small, consistent with Remark~3.}

\subsection{Real-data-based {nonparametric} simulations} \label{sec:application}
This section illustrates the proposed methods using simulations based on a widely used benchmark data from the Right Heart Catheterization (RHC) observational study \citep{Connors96}, which was conducted to evaluate the causal effect of the diagnostic cardiological procedure of RHC (binary treatment) among hospitalized critically ill patients. The purpose of our simulation is not to re-design an existing study {or conduct a \emph{post-hoc} power analysis}; instead, we want to assess the performance of the proposed methods in a real world setting where we do not know the true data generating model, presenting a more realistic and challenging setting. The data contain 5735 patients, each with 51 covariates (20 continuous, 25 binary, 6 categorical). The outcome is the binary survival status at 30 days after admission, with $Y=1$ meaning death. In total, 2184 ($38.1\%$) of the patients received the treatment and the rest did not. The death rate is 38.0\% and 30.6\% in the treatment and control group, respectively. 

The simulation is outlined as follows. First, we calculate the summary parameters $(r, \phi, \rho^2)$ and the effect size $\widetilde{\tau}^{\ATE}$ from the real data. Second, we take these values as the only inputs of a hypothetical power analysis {for a new observational study with the same goal on a similar population}. We apply the proposed formula in Theorem \ref{thm:sample-size-formula} with these inputs to compute the minimal sample size $N$ needed to achieve a certain level of power in detecting the pre-fixed effect size $\widetilde{\tau}^{\ATE}$. Third, we reverse the process to estimate the power using repeated bootstrapping samples of size $N$ randomly drawn from the RHC data. The closer the estimated power is to the nominal power (e.g. 0.8), the more accurate our method is. {This simulation is nonparametric because we do not assume any outcome model.  } 

Following the above outline, from the real data we estimate the propensity scores using a logistic model with the main effects of each covariate. Based on these estimated scores, we numerically compute the true overlap coefficient $\phi$ defined in \eqref{eq:overlap-def} to be 0.835. We also used the proposed normal approximation in Section \ref{sec:dist-ps}, $W_e \sim \mathsf{N}(-0.701, 2.189)$, and estimate the overlap coefficient to be 0.822, which is close to the true value, showing the precision of our proposed Beta approximation. The H\'{a}jek estimate of the standardized effect size $\widetilde{\tau}^{\ATE}$ is 0.14. The treatment proportion is $r=0.381$ and $\rho^2=0.000$. With these inputs alone, we apply formula \eqref{eq:ss-formula} to calculate the minimal sample size to achieve 0.8 power to be 3625. We then reverse the process to calculate the actual power under a sample of 3625 units. Specifically, we generate $B = 1000$ replicates, each with 3625 rows randomly drawn from the original data with replacement. We calculate the H\'{a}jek estimate of $\tau^{\ATE}$ and the associated $95\%$ confidence interval in each replicate. Among the 1000 replicates, 924 intervals do not cover zero, giving an estimated power of 0.924 with 95\% confidence interval (0.908 -- 0.940), which is larger than the nominal power of 0.8. Potential reasons for the discrepancy include: (i) we used the estimated instead of the true propensity scores, and (ii) we treated the binary outcome as continuous. We also computed the sample size from the two-sample $z$-test, which is 1567, leading to a vastly underpowered estimate with the power of only 0.55. Given that any real study would have data attrition due to missing data and extreme weights, a conservative (overpowered) estimate of the sample size is arguably more desirable than the opposite.

\begin{figure}
    \centering
    \includegraphics[width=0.85\linewidth]{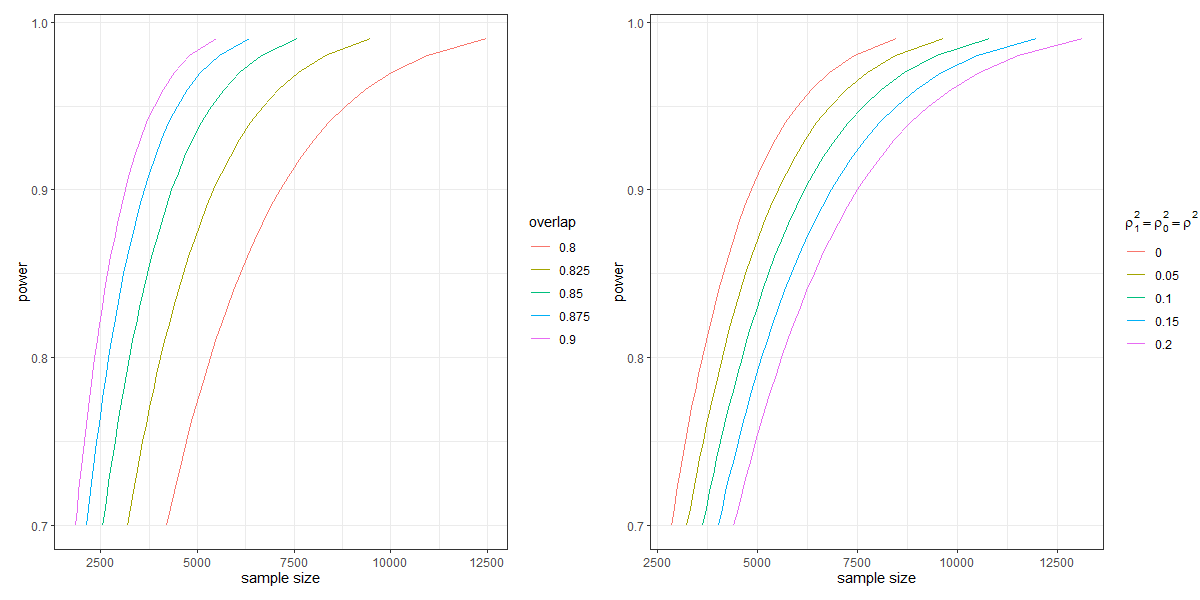}
    \caption{Power and sample size curves of an emulated  Right Heart Catheterization  study with an effect size $\widetilde{\tau}=0.14$ under different overlap coefficient $\phi$ (left) and the correlation $\rho=\rho_1=\rho_0$ (right).}
    \label{fig:sensitivity}
\end{figure}

Figure \ref{fig:sensitivity} shows the power curves of $\tau^{\ATE}$ given different values of the overlap coefficient $\phi$ and the correlation $\rho^2$ while fixing other summary inputs as those calculated from the RHC data. It shows that, to achieve the same power, the minimal sample size would increase as either the overlap $\phi$ decreases or the correlation $\rho$ increases, matching the theoretical result in Corollary \ref{thm:ss-monotone}. The $R^2$ value is 0.21, which serves as the upper bound of $\rho^2$. 

We repeat the process for the ATO and ATT estimand, respectively, and the results suggest the proposed formula lead to reliable power calculation in a wide range of scenarios. Details are relegated to Appendix {C.4}. 

\section{Discussion} \label{sec:conclusion}
This paper establishes the theoretical foundation of power analysis for causal inference with observational data. We show that, compared with the power calculation of randomized trials, users only need to specify two additional parameters for emulated trials: an overlap coefficient $\phi$ that quantifies the confounder-treatment association and a confounding coefficient $\rho$ that quantifies the confounder-outcome association. We propose interpretable and easily computable statistics to characterize these two parameters. {Our method requires a logistic model for the propensity scores and a homoscedastic restricted mean outcome model with a homogeneous treatment effect}, but it bypasses distributional assumptions on multivariate covariates. In practice, investigators can set these parameters based on a pilot study or similar previous studies, or conduct sensitivity analysis of these parameters based on domain knowledge. 

Some authors \citep[e.g.][]{hernan2022causal} questioned the use of power analysis in existing observational studies; they argued that the primary consideration in observational studies is to reduce confounding bias rather than to achieve sufficient power. We agree that de-confounding is central in observational studies, but that  does not preclude the critical role of power analysis in designing observational studies, which are usually subject to cost and time constraints. For example, in prospective studies of rare or severe diseases, recruiting patients is time-consuming, and thus investigators must determine the minimal sample size to achieve a reliable estimate before conducting the study. In this type of cohort studies, ideally investigators could recruit participants in batches, with the first batch serving as a pilot study, which provides reliable estimate of the key inputs $(\phi,\rho)$ and enables a more accurate calculation of the sample size needed for subsequent batches. With existing observational data, power analysis is still crucial in helping researchers understand the scope of data required and guide database selection in the design stage. An example concerns retrospective commercial population health data, such as the Truveta electronic health record data. These data sources usually charge users costs proportional to the data size and computational load. With limited resources, a sample size calculation is crucial for designing studies based on such data sources. Moreover, the same data source (e.g. disease registries or  electronic health records of a population) is often used for investigating multiple hypotheses, e.g. effects on secondary outcomes in clinical research. In these cases, power analysis can help researchers decide how aggressive they can be with multiplicity correction or if they have to whittle down the primary objective.

A power analysis must be conducted in the design stage of a study. Once the study data are fixed and the analysis has been carried out, it is not meaningful and in fact is misleading to perform a \emph{post-hoc} power analysis on the fixed data, e.g. using the bootstrapping procedure in the RHC simulation in Section \ref{sec:application}. Such simulations cannot capture all the uncertainties in estimating the causal effects because of the conditioning on the observed sample. Moreover, the observed data may not satisfy the model assumptions underlying the sample size formula that the \emph{post-hoc} power analysis implicitly assumes, creating a design-analysis mismatch.

We anchor the power calculation on the H\'{a}jek estimator, which is less efficient than some alternatives, such as the doubly robust estimators that attain the semiparametric efficiency \citep{Robins1994doublerobust, tsiatis2006semiparam}. The asymptotic variance of these alternatives, however, often depends on outcome models whose form cannot be anticipated at the design stage. A power calculation built on such an estimator would presuppose a specific outcome model, risking a design-analysis mismatch if that model is not used in analysis. Moreover, the efficiency gains typically rely on correct model specification, which is far more difficult to verify before data collection than after. Sample size formulas based on the H\'{a}jek estimator therefore protect against underpowered studies, since they does not presuppose the successful use of any efficiency tool. The proposed framework requires only a logistic propensity model and a semiparametric restricted mean outcome model, imposing minimal model assumptions. At the analysis stage, investigators may still apply more efficient estimators as a safeguard against null findings or to support testing of secondary hypotheses. Extending the proposed framework to derive sample size formulas under stronger assumptions tied to specific efficient estimators is a natural direction for future work.

There are several extensions of the proposed method in observational studies, including repeated outcomes, time-to-event outcomes, and multiple treatments. Propensity score weighting estimators for causal effects are available in all these settings. Therefore, we can readily adapt the general structure of power calculation developed in this paper. For example, for studies with repeated outcomes, a common estimand is the time-averaged difference; we can analyze the variance of an inverse probability weighted marginal outcome model within-subject correlation between outcomes. For time-to-event outcomes, a common estimand is the marginal causal hazard ratio from the Cox proportional hazard model. \cite{yang2026sample} derived the associated analytical sample size formula for both randomized trials and observational studies based on the asymptotic variance of the robust sandwich estimator of the inverse probability weighted Cox model.

\section*{Acknowledgements}
The authors thank the associate editor, three anonymous reviewers,  Xiaoxiao Zhou, Jay Lusk, and Laine Thomas for constructive comments. This paper is dedicated to the memory of the late Sayan Mukherjee, who taught FL basics of power analysis and beyond.   

\bibliographystyle{apalike}
\bibliography{power}  

\newpage
\begin{center}
    \Large\bf Appendix
\end{center}

\begin{appendix}

\section{Proof of Propositions and Theorems}

\subsection{Proof of Equation (5)}
Rewrite $\widehat{\tau}_N = \tau + \epsilon \sqrt{V_0/N}$, where $\epsilon \sim \mathsf{N}(0, 1)$. Then the power of test is given by
\begin{align}
    \Pr\left(\tau + \epsilon\sqrt{\frac{V_0}{N}} > z_{1-\alpha}\sqrt{\frac{V}{N}}\right) &= \Pr\left(\epsilon > z_{1-\alpha} \sqrt{\frac{V}{V_0}} - \tau \sqrt{\frac{N}{V_0}}\right) \\
    &= 1 - \Phi\left(\frac{z_{1-\alpha} \sqrt{V} - \tau \sqrt{N}}{\sqrt{V_0}}\right).
\end{align}
Equalizing this power with $\beta$, we obtain the sample size formula 
\begin{equation}
    N = \frac{\left(z_{1-\alpha} \sqrt{V} - z_{1-\beta} \sqrt{V_0}\right)^2}{\tau^2}.
\end{equation}
Furthermore, when $\alpha < 0.5 < \beta$, $z_{1-\alpha} > 0$, $z_{1-\beta} < 0$, and hence we have 
\begin{equation}
    N \leq \frac{V}{\tau^2}\left(z_{1-\alpha} - z_{1-\beta}\right)^2 = \frac{V}{\tau^2}\left(z_{1-\alpha} + z_{\beta}\right)^2.
\end{equation}

\subsection{Special case of the Lyapunov Central Limit Theorem}

For completeness, we first present the general form of the Lyapunov Central Limit Theorem (CLT).
\begin{theorem}[Lyapunov CLT]
    Suppose $\{Z_1, \dots, Z_n, \dots\}$ is a sequence of independent random variables, each with finite expected value $\mu_i$ and variance $\sigma_i^2$. Define $s_n^2 = \sum_{i=1}^n \sigma_i^2$. If for some $\delta > 0$, the Lyapunov’s condition 
    \begin{equation}
        \lim_{n\rightarrow \infty} \frac{1}{s_n^{2+\delta}}\sum_{i=1}^n \mathbb{E}\left[|Z_i - \mu_i|^{2 + \delta}\right] = 0
    \end{equation}
    is satisfied, then 
    \begin{equation}
        \frac{1}{s_n} \sum_{i=1}^n (Z_i - \mu_i) \Rightarrow \mathsf{N}(0, 1).
    \end{equation}
\end{theorem}

As a special case, let $Z_i = \beta_i X_i$, where $\{X_1, \cdots, X_n, \cdots\}$ is a sequence of independent random variables with expectation 0 and variance 1. Then $Z_i$ has expectation 0 and variance $\beta_i^2$. Assume (1) $\mathbb{E}[X_i^4] \leq B < \infty$ and (2) $M_n \equiv \frac{\max_{1\leq i \leq n} \beta_i^2}{\sum_{i=1}^n \beta_i^2} = o(1)$, then the Lyapunov's condition as satisfied (by taking $\delta = 2$) because
\begin{align*}
    \lim_{n\rightarrow \infty} \frac{1}{(\sum_{i=1}^n \beta_i^2)^2}\sum_{i=1}^n \beta_i^4 \mathbb{E}[X_i^4] &\leq B \lim_{n\rightarrow \infty} \frac{\sum_{i=1}^n \beta_i^4}{(\sum_{i=1}^n \beta_i^2)^2} \\
    &\leq B \lim_{n\rightarrow \infty} \frac{\max_{1\leq i \leq n} \beta_i^2 \sum_{i=1}^n \beta_i^2}{(\sum_{i=1}^n \beta_i^2)^2} \\
    &= B\lim_{n\rightarrow \infty} M_n \\
    &= 0.
\end{align*}
Therefore, $\sum_{i=1}^n \beta_i X_i$ is approximately a normal distribution with mean 0 and variance $\sum_{i=1}^n \beta_i^2$ when $n$ is large.

\subsection{Proof of Proposition 1(i)}

Suppose $U \sim \mathsf{Beta}(a, b)$ and $Z \sim \mathsf{Bin}(U)$. Then, 
\begin{equation}
    \mathbb{E}[Z] = \mathbb{E}[\mathbb{E}[Z \mid U]] = \mathbb{E}[U] = \frac{a}{a+b}.
\end{equation}

Furthermore, define $f_z(u) = \Pr(U = u \mid Z = z)$ for $z \in \{0, 1\}$. By Bayes' rule, 
\begin{equation}
    f_z(u) = \Pr(U = u\mid Z = z) = \frac{\Pr(U = u) \Pr(Z = z \mid U = u)}{\Pr(Z = z)}.
\end{equation}
Therefore,
\begin{align*}
    f_1(u) &= \frac{\Pr(U = u)u}{a / (a + b)} = \frac{\Gamma(a + b)}{\Gamma(a)\Gamma(b)}\frac{u^a (1 - u)^{b - 1}}{a / (a + b)} = \frac{\Gamma(a + b + 1)}{\Gamma(a + 1) \Gamma(b)} u^a (1 - u)^{b - 1}, \\
    f_0(u) &= \frac{\Pr(U = u)(1 - u)}{b / (a + b)} = \frac{\Gamma(a + b)}{\Gamma(a)\Gamma(b)}\frac{u^{a - 1} (1 - u)^b}{b / (a + b)} = \frac{\Gamma(a + b + 1)}{\Gamma(a) \Gamma(b + 1)} u^{a - 1} (1 - u)^{b}.
\end{align*}
Consequently,
\begin{align*}
    \phi &= \int_0^1 \sqrt{\frac{\Gamma(a+b+1)^2}{\Gamma(a+1)\Gamma(b)\Gamma(a)\Gamma(b+1)}} u^{a - 0.5}(1-u)^{b-0.5}\,\mathrm{d} u \\
    &= \frac{\Gamma(a+b+1)}{\sqrt{a}\Gamma(a)\sqrt{b}\Gamma(b)}\cdot \frac{\Gamma(a + 0.5)\Gamma(b + 0.5)}{\Gamma(a + b + 1)} \\
    &= \frac{\Gamma(a + 0.5)}{\sqrt{a}\Gamma(a)}\frac{\Gamma(b + 0.5)}{\sqrt{b}\Gamma(b)}.
\end{align*}

\subsection{Proof of Proposition 1(ii)} \label{app:proof-propii}

Denote $\phi$ by $\phi(a,b)$ to indicate that $\phi$ is a function of $a$ and $b$, and factorize it as $\phi(a,b)= \tilde{\phi}(a) \tilde{\phi}(b)$, where $\tilde{\phi}(x)=\Gamma(x + 0.5)/\{\sqrt{x}\Gamma(x)\}$.

To show $\phi(a,b)$ increases monotonically in $a$ (and $b$), it suffices to show that the partial derivative of $\log \phi(a,b) = \log \tilde{\phi}(a) + \log \tilde{\phi}(b)$ with respect to $a$ (and $b$) is positive. By seeing that $\log \tilde{\phi}(x) = \log \Gamma(x+0.5) - \log \Gamma(x) - \log \sqrt{x}$, we have

\begin{equation}
\label{eq:derivative-phi}
    \frac{\partial \log \phi(a,b)}{\partial a} = \frac{d \log \tilde{\phi}(a)}{d a} = \psi(a+\frac{1}{2}) - \psi(a) - \frac{1}{2a},
\end{equation}
where $\psi(x) = d \log \Gamma(x)/d x$ is the digamma function.

Leveraging \cite{abramowitz1965handbook} Equation (6.3.22), we have
\begin{equation}
\label{eq:digamma-int}
        \psi(x) - \psi(y) = \int_0^1 \frac{t^{y-1} - t^{x-1}}{1 - t}\,\mathrm{d}t, \quad \forall x, y > 0.
\end{equation}
Plugging into Equation~\eqref{eq:derivative-phi} yields
\begin{equation}
    \psi(a+\frac{1}{2}) - \psi(a) = \int_0^1 \frac{t^{a-1} - t^{a-1/2}}{1 - t} \,\mathrm{d}t
    %&= \int_0^1 \frac{ t^{a-1} (1 - \sqrt{t})}{1 - t} \,\mathrm{d}t \ - \frac{1}{2a} \\
    = \int_0^1 \frac{ t^{a-1}}{1 + \sqrt{t}} \,\mathrm{d}t
    > \int_0^1 \frac{ t^{a-1}}{2} \,\mathrm{d}t
    = \frac{1}{2a},
\end{equation}
which leads to
\begin{equation}
\label{eq:derivative-phi-pos}
    \frac{\partial \log \phi(a,b)}{\partial a} > 0.
\end{equation}
Therefore, $\log \phi(a,b)$ increases monotonically in $a$ and so does $\phi(a,b)$. By symmetry, $\phi(a,b)$ also increases monotonically in $b$.

\subsection{Proof of Proposition 1(iii)} \label{app:proof-propiii}

To see the 1-1 mapping between $(r, \phi)\in(0,1)^2$ and $(a,b)\in (0, +\infty)^2$, it suffices to provide an injective and surjective mapping from $(a,b)$ to $(r, \phi)$. Define the mapping $h : (a,b) \to (r, \phi)$ as
\begin{equation}
    h: (a,b) \to \left( r = \frac{a}{a+b}, \ \phi = \tilde{\phi}(a) \tilde{\phi}(b) \right),
\end{equation}
where $\tilde{\phi}(x)$ is defined in Appendix~\ref{app:proof-propii}.

Before showing that $h$ is injective and surjective, we first establish the properties of the restricted univariate mapping from $a$ to $\phi$ for arbitrarily fixed $r \in (0,1)$ induced by $h$, denoted as $\phi_h(a;r)$. For fixed $r$, we have $b=a(1-r)/r$, and $h$ maps $(a, b)$ to $(r, \phi_h(a;r))$, where $\phi_h(a;r)$ is defined as
\begin{equation}
    \phi_h(a;r) = \tilde{\phi}(a) \tilde{\phi}(b)= \tilde{\phi}(a) \tilde{\phi} \left( a(1-r)/r \right).
\end{equation}

First, $\phi_h(a;r)$ increases monotonically in $a$, because
\begin{equation}
    \frac{d \log \phi_h(a;r)}{d a} = \frac{d \log \tilde{\phi}(a)}{d a} + \frac{d \log \tilde{\phi}(b)}{d a} = \frac{d \log \tilde{\phi}(a)}{d a} + \frac{d \log \tilde{\phi}(b)}{d b} \frac{db}{da} > 0,
\end{equation}
by Equation~\eqref{eq:derivative-phi}, Equation~\eqref{eq:derivative-phi-pos}, and $db/da=(1-r)/r>0$.

Second, the image of $\phi_h(a;r)$ is $(0,1)$, i.e., $\phi_h(a;r)$ maps $a \in (0, +\infty)$ onto the domain of $\phi$ excluding $0$ and $1$. Evaluate the limit of $\tilde{\phi}(x)$ at $x \to 0^+$ and $x \to +\infty$ respectively, we have
\begin{equation}
    \lim_{x \to 0^+} \tilde{\phi}(x) = \lim_{x \to 0^+} \frac{\Gamma(x + 0.5)} {\sqrt{x}\Gamma(x)} = \lim_{x \to 0^+} \sqrt{x} \frac{\Gamma(x+0.5)}{\Gamma(x+1)} = 0 \times \frac{\sqrt{\pi}}{1} = 0,
\end{equation}
by using $\Gamma(x)=\Gamma(x+1)/x$, and
\begin{equation}
    \lim_{x \to +\infty} \tilde{\phi}(x) = \lim_{x \to +\infty} \frac{\Gamma(x + 0.5)} {\sqrt{x}\Gamma(x)} = 1,
\end{equation}
by Equation (12) in \cite{Wendel1948NoteOT}. Since both limits are finite and $a \to 0^+$ (or $+\infty$) implies $b \to 0^+$ (or $+\infty$) respectively, therefore
\begin{equation}
    \lim_{a \to 0^+} \phi_h(a;r) = 0, \quad \lim_{a \to +\infty} \phi_h(a;r) = 1.
\end{equation}
Given that $\phi_h(a;r)$ is continuous and monotonically increasing in $a$, by intermediate value theorem, for any $\phi^* \in (0,1)$ there exists a unique $a^*$ such that $\phi_h(a^*;r)=\phi^*$. Therefore, the image of $\phi_h(a;r)$ is $(0,1)$.

Now, we show $h$ is injective and surjective. For injectivity, suppose that $h$ maps $(a_1, b_1)$ and $(a_2, b_2)$ to the same $(r, \phi)$. Therefore, $\phi_h(a_1;r) = \phi_h(a_2;r) = \phi$. Since $\phi_h(a;r)$ is monotonic in $a$, so $a_1=a_2$, hence $b_1=a_1(1-r)/r=a_2(1-r)/r=b_2$, establishing injectivity. 

For surjectivity, pick an arbitrary $(r^*, \phi^*) \in (0,1)^2$. It induces $\phi_h(a;r^*)$ which has the image of $(0,1)$, so there exists $a^*$ such that $\phi_h(a^*;r^*) =\phi^*$. Define $b^*=a^*(1-r^*)/r^*$, then $h$ maps $(a^*, b^*)$ to $(r^*, \phi^*)$, establishing surjectivity.

\subsection{Proof of Equation (12)} \label{app:solve-params}

Recall $e(X) = \expit(W_e)$ and $W_e \sim \mathsf{N}(\mu_e, \sigma_e^2)$ given in Section 3.2 and $Y(z) = aW_e + z\tau + \epsilon$ given in Section 3.3, where $\epsilon$ is the compound error and $\bE[\epsilon]=\mu$ and $\bV[\epsilon]=\sigma_y^2$. For clarity of the proof, we write out $\epsilon = W_e^\perp + \epsilon_{ori}$, where $\epsilon_{ori}$ is the original error term defined in the restricted mean model (RMM).

By the definition of RMM, $\epsilon_{ori}$ is uncorrelated with any measurable function of $X$, including $W_e$, and therefore $\Cov(W_e,\epsilon)=\Cov(W_e,W_e^\perp)+\Cov(W_e,\epsilon_{ori})=0$. Hence we have the following results:
\begin{align}
    S^2 &\equiv \mathbb{V}[Y(0)] = \mathbb{V}[aW_e + \epsilon] = a^2 \sigma_e^2 + \sigma_y^2; \label{eq:v_z}\\
    \rho &\equiv \mathrm{cor}[Y(0), W_e] = \mathrm{cor}[a W_e + \epsilon, W_e] = a\sqrt{\sigma_e^2 / S^2 }. \label{eq:rho_z}
\end{align}
By \eqref{eq:rho_z},
$a = \rho \sqrt{{S^2}/\sigma_e^2}$,
which, plugged into \eqref{eq:v_z}, yields
\begin{equation}
     \sigma_y^2 = (1 - \rho^2) S^2.
\end{equation}

\vspace{-12pt}

\subsection{Proof of the analytical form of variance $V$ in Theorem 2}\label{app:proof-thm3.2}

Recall that 
\begin{equation}
    V = \mathbb{E}\left[\frac{\{Y(1) - \mathbb{E}[Y(1)]\}^2}{e(X)}\right] + \mathbb{E}\left[\frac{\{Y(0) - \mathbb{E}[Y(0)]\}^2}{1 - e(X)}\right],
\end{equation}
where $e(X)$ and $Y(z)$ are given in Appendix~\ref{app:solve-params}. We first state all necessary results regarding $\epsilon$, $\epsilon_{ori}$, $W_e$, and $W_e^\perp$ in Lemma~\ref{lem:rmm}.
\begin{lemma}\label{lem:rmm}
Under the joint normality of $(f(X),W_e)$ and the RMM Model:
\begin{enumerate}[label=(\arabic*), itemindent=0.5in]
\item $\bE[\epsilon_{ori}|W_e]=0$
\item $\mu=\bE[\epsilon]=\bE[\epsilon | W_e]=\bE[W_e^{\perp}]$
\item $\bE[\epsilon_{ori} | W_e, W_e^{\perp}]=0$
\item $\Cov(W_e^\perp, \epsilon_{ori})=0$
\end{enumerate}
\end{lemma}
Lemma~\ref{lem:rmm} holds because: (1) $\bE[\epsilon_{ori}|X]=0$ and the law of iterated expectation, see Section 2.2.4 of \cite{woodridge2010EconBook}, that $\bE[\epsilon_{ori}|W_e] = \bE \left\{ \bE[\epsilon_{ori}|X] \mid W_e \right\} = 0$. (2) $W_e$ and $W_e^\perp$ are independent, and $\bE[\epsilon | W_e] = \bE[W_e^{\perp} | W_e] + \bE[\epsilon_{ori} | W_e] = \bE[W_e^{\perp}]$. (3) $\bE[\epsilon_{ori} | W_e, W_e^{\perp}] = \bE[\epsilon_{ori} | W_e, f(X)] = \bE \left\{ \bE[\epsilon_{ori}|X] \mid W_e, f(X) \right\} =0$, see that any one of $(f(X),W_e,W_e^{\perp})$ is determined by another two. (4) $W_e^{\perp}=f(X)-aW_e$ is a function of $X$ and therefore uncorrelated with $\epsilon_{ori}$ by the definition of RMM.

We derive the first term in $V$, and similar result holds for the second term by symmetry.
\begin{align}
    & \mathbb{E} \left[\frac{\{Y(1) - \mathbb{E}[Y(1)]\}^2}{e(X)}\right] = 
    \mathbb{E} \left[\frac{ \{ a(W_e - \mu_e) + (\epsilon - \mu) \}^2 }{\expit (W_e)}\right] \\
    & = a^2\mathbb{E} \left[\frac{ (W_e - \mu_e)^2 }{\expit (W_e)}\right] + \mathbb{E} \left[\frac{ (\epsilon - \mu)^2 }{\expit (W_e)}\right] + 2a\mathbb{E} \left[\frac{ (W_e - \mu_e)(\epsilon - \mu)  }{\expit (W_e)}\right] \\
    & := A + B + C
\end{align}
Term $A$ only depends on the propensity score model but not the outcome model, while $B$ and $C$ involve both models. We first derive $B$ and $C$.
\begin{align}\label{eq:termB}
    B & = \mathbb{E} \left\{ \mathbb{E} \left[ \frac{ (\epsilon - \mu)^2 }{\expit (W_e)} \Bigg| W_e \right] \right\} = \mathbb{E} \left\{ \frac{ \bE[(\epsilon - \mu)^2 \mid W_e] }{\expit (W_e)} \right\} = \mathbb{E} \left\{ \frac{ \bV[\epsilon \mid W_e] }{\expit (W_e)} \right\},
\end{align}
and for $\bV[\epsilon \mid W_e]$ we have
\begin{align}
    \bV[\epsilon \mid W_e] &= \bV[W_e^\perp | W_e] + \bV[\epsilon_{ori}|W_e] + 2\Cov(W_e^\perp, \epsilon_{ori} | W_e).
\end{align}
For these three pieces: $\bV[W_e^\perp | W_e] = \bV[W_e^\perp]$ because $W_e$ is independent of $W_e^\perp$; $\bV[\epsilon_{ori}|W_e] = \bV[\epsilon_{ori}]$ because of the homoskedasticity assumption that $\bV[\epsilon_{ori}|e(X)] = \bV[\epsilon_{ori}]$, and $W_e$ is a deterministic function of $e(X)$; and for the last piece,
\begin{align}
    \Cov(W_e^\perp, \epsilon_{ori} | W_e) &= \bE[W_e^\perp \cdot \epsilon_{ori}|W_e] - \bE[W_e^\perp|W_e] \bE[\epsilon_{ori}|W_e] \\
    &= \bE\left\{W_e^\perp \cdot \bE[\epsilon_{ori}|W_e, W_e^\perp] \ \big| W_e \right\} - 0, \quad \text{by Lemma~\ref{lem:rmm} (1)} \\
    & = \bE\left\{W_e^\perp \cdot 0  \ \big| W_e \right\} =0, \quad \text{by Lemma~\ref{lem:rmm} (3).}
\end{align}
Therefore, $\bV[\epsilon \mid W_e]$ is invariant to $W_e$ by seeing that
\begin{equation}
    \bV[\epsilon \mid W_e] = \bV[W_e^\perp] + \bV[\epsilon_{ori}] + 0 = \bV[\epsilon] = \sigma_y^2,
\end{equation}
and consequently
\begin{equation}
    B = \mathbb{E} \left\{ \frac{ \bV[\epsilon \mid W_e] }{\expit (W_e)} \right\} = \sigma_y^2 \cdot \mathbb{E}\left[\frac{1}{\expit (W_e)}\right]
\end{equation}

Finally, for $C$,
\begin{align}
    C & = 2a \cdot \mathbb{E} \left\{ \frac{W_e - \mu_e}{\expit (W_e)} \bE[ \epsilon - \mu \mid W_e  ] \right\} \\
    &= 2a \cdot \mathbb{E} \left\{ \frac{W_e - \mu_e}{\expit (W_e)} (\bE[\epsilon|W_e] - \bE[W_e^\perp] ) \right\}, \quad \text{by Lemma~\ref{lem:rmm} (2)} \\ 
    &= 2a \cdot \mathbb{E} \left\{ \frac{W_e - \mu_e}{\expit (W_e)} \cdot 0 \right\} = 0.
\end{align}

Gathering $A$, $B$, and $C$ above,
\begin{align}
    \mathbb{E}\left[\frac{\{Y(1) - \mathbb{E}[Y(1)]\}^2}{e(X)}\right] = a^2 \cdot \mathbb{E}\left[\frac{(W_e - \mu_e)^2}{\expit (W_e)}\right] + \sigma_y^2 \cdot \mathbb{E}\left[\frac{1}{\expit (W_e)}\right],
\end{align}
which is fully determined under $a^2$, $\sigma_y^2$, and the distribution of $W_e$, as calculated below.

\begin{align}
    \mathbb{E}\left[\frac{1}{\expit (W_e)}\right] &= \int \left(1 + e^{-w}\right) \frac{1}{\sqrt{2\pi \sigma_e^2}}\exp\left(-\frac{(w - \mu_e)^2}{2\sigma_e^2}\right)\,\mathrm{d}w \\
    &= \int \left(1 + e^{-u - \mu_e}\right) \frac{1}{\sqrt{2\pi \sigma_e^2}}\exp\left(-\frac{u^2}{2\sigma_e^2}\right)\,\mathrm{d}u\\
    &= 1 + \exp\left(-\mu_e + \frac{1}{2}\sigma_e^2\right),
\end{align}

\begin{align}
    \mathbb{E}\left[\frac{(W_e - \mu_e)^2}{\expit (W_e)}\right] &= \int (w - \mu_e)^2 \left(1 + e^{-w}\right) \frac{1}{\sqrt{2\pi \sigma_e^2}}\exp\left(-\frac{(w - \mu_e)^2}{2\sigma_e^2}\right)\,\mathrm{d}w \\
    &= \int u^2 \left(1 + e^{-u - \mu_e}\right) \frac{1}{\sqrt{2\pi \sigma_e^2}}\exp\left(-\frac{u^2}{2\sigma_e^2}\right)\,\mathrm{d}u\\
    &= \sigma_e^2 + (\sigma_e^4 + \sigma_e^2)\exp\left(-\mu_e + \frac{1}{2}\sigma_e^2\right).
\end{align}

Therefore, 
\begin{equation}\label{eq:V1}
    \mathbb{E}\left[\frac{\{Y(1) - \mathbb{E}[Y(1)]\}^2}{e(X)}\right] = a^2\sigma_e^2 + \sigma_y^2 + \left[a^2\sigma_e^2(\sigma_e^2 + 1) + \sigma_y^2\right]\exp\left(-\mu_e + \frac{1}{2}\sigma_e^2\right).
\end{equation}

Similarly, 
\begin{equation}\label{eq:V0}
    \mathbb{E}\left[\frac{\{Y(0) - \mathbb{E}[Y(0)]\}^2}{1 - e(X)}\right] = a^2\sigma_e^2 + \sigma_y^2 + \left[a^2\sigma_e^2(\sigma_e^2 + 1) + \sigma_y^2\right]\exp\left(\mu_e + \frac{1}{2}\sigma_e^2\right).
\end{equation}

Adding \eqref{eq:V1} and \eqref{eq:V0}, we have
\begin{eqnarray}
    V = 2(a^2 \sigma_e^2 + \sigma_y^2) + 2\left[a^2\sigma_e^2(\sigma_e^2 + 1) + \sigma_y^2\right]\exp\left(\frac{\sigma_e^2}{2}\right)\cosh(\mu_e).
\end{eqnarray}

Plugging in the expression for $a$ and $\sigma_y^2$,
\begin{equation}
    V = 2S^2\left\{1 + (\rho^2\sigma_e^2 + 1)\exp\left(\frac{\sigma_e^2}{2}\right)\cosh(\mu_e)\right\}.
\end{equation}

\vspace{-12pt}

\subsection{Proof of the asymptotic variance of the general H\'{a}jek WATE estimator}

Let $$\xi_1^* = \frac{\mathbb{E}[h(X_i; \alpha^*) Y_i(1)]}{\mathbb{E}[h(X_i; \alpha^*)]}, \quad \xi_0^* = \frac{\mathbb{E}[h(X_i; \alpha^*) Y_i(0)]}{\mathbb{E}[h(X_i; \alpha^*)]}$$ be the expected potential outcomes given the tilting function $h(x)$ and let $\alpha^*$ be the true parameter value for the propensity score model $e(X; \alpha)$.
Let $$\hat{\xi}_1 = \frac{\sum_{i=1}^N Z_i Y_i \hat{w}_1(X_i)}{\sum_{i=1}^N Z_i \hat{w}_1(X_i)}, \quad \hat{\xi}_0 = \frac{\sum_{i=1}^N (1 - Z_i) Y_i \hat{w}_0(X_i)}{\sum_{i=1}^N (1 - Z_i) \hat{w}_0(X_i)},$$ and $$\hat{\tau}_w = \hat{\xi}_1 - \hat{\xi}_0.$$

Then $(\hat{\xi}_1, \hat{\xi}_0, \hat{\alpha})$ jointly solves
$$
0 = \frac{1}{N}\sum_{i=1}^N \phi(Y_i(1), Y_i(0), x_i, Z_i; \xi_1, \xi_0, \alpha) = \frac{1}{N}\sum_{i=1}^N\begin{pmatrix}
  \phi_1(Y_i(1), x_i, Z_i; \xi_1, \alpha) \\
  \phi_0(Y_i(0), x_i, Z_i; \xi_0, \alpha) \\
  \psi_\alpha(x_i, Z_i; \alpha),
\end{pmatrix}$$
where 
\begin{equation}
  \phi = \begin{pmatrix}
    \phi_1 \\ \phi_0 \\ \psi_\alpha
  \end{pmatrix}
  \quad \text{with} \quad
  \begin{aligned}
    \phi_1 &= Z_i\cdot w_1(X_i; \alpha)\cdot \{Y_i(1) - \xi_1\}, \\
    \phi_0 &= (1 - Z_i) \cdot  w_0(X_i; \alpha) \cdot \{Y_i(0) - \xi_0\}, \\
    \psi_\alpha &= e'(X_i; \alpha )\{Z_i - e(X_i; \alpha)\}.
  \end{aligned}
\end{equation}
It is straightforward to verify that $\mathbb{E}[\phi(\xi_1^*, \xi_0^*, \alpha^*)] = 0$. The theory of M-estimation states that 
\begin{equation}
  \sqrt{N}\begin{pmatrix}
    \hat{\xi}_1^ - \xi_1^* \\ \hat{\xi}_0^ - \xi_0^* \\ \hat{\alpha} - \alpha^*
  \end{pmatrix} \stackrel{d}{\rightarrow} \mathcal{N}(0, (A^*)^{-1}B^* (A^*)^{-\top}),
\end{equation}
where $A^*$ and $B^*$ are the values of $-\mathbb{E}\left(\frac{\partial \phi}{\partial (\xi_1, \xi_0, \alpha)}\right)$ and $\mathbb{E}[\phi \phi^\top]$ evaluated at $(\xi_1^*, \xi_0^*, \alpha^*)$ respectively.

\textbf{Calculation of $A^*$.} 

By definition, we have 
\begin{equation}
  \frac{\partial \phi}{\partial (\xi_1, \xi_0, \alpha)}\big|_{(\xi_1^*, \xi_0^*, \alpha^*)} = \begin{pmatrix}
    \frac{\partial \phi_1^*}{\partial \xi_1} & 0 & \frac{\partial \phi_1^*}{\partial \alpha^\top} \\
    0 & \frac{\partial \phi_0^*}{\partial \xi_0} & \frac{\partial \phi_0^*}{\partial \alpha^\top} \\
    0 & 0 & \frac{\partial \psi_\alpha^*}{\partial \alpha^\top},
  \end{pmatrix}
\end{equation}
where 
$$\begin{aligned}
  \frac{\partial \phi_1^*}{\partial \xi_1} &= -Z_i\cdot w_1(X_i; \alpha^*), \\
  \frac{\partial \phi_0^*}{\partial \xi_0} &= -(1 - Z_i)\cdot w_0(X_i; \alpha^*), \\
  \frac{\partial \phi_1^*}{\partial \alpha} &= Z_i\cdot \{Y_i(1) - \xi_1^*\} \cdot w_1'(X_i;\alpha^*), \\
  \frac{\partial \phi_0^*}{\partial \alpha} &= (1 - Z_i)\cdot \{Y_i(0) - \xi_0^*\} \cdot w_0'(X_i; \alpha^*), \\
  \frac{\partial \psi_\alpha^*}{\partial \alpha} &= e''(X_i; \alpha^*)(Z_i - e(X_i; \alpha^*)) - \left[e'(X_i; \alpha^*)\right]\left[e'(X_i; \alpha^*)\right]^\top.
\end{aligned}$$
Accordingly, we have 
$$A^* = \begin{pmatrix}
  a_{11} & 0 & a_{13} \\
  0 & a_{22} & a_{23} \\
  0 & 0 & a_{33}
\end{pmatrix},$$
where 
$
%\begin{aligned}
  a_{11} = \mathbb{E}\left[Z_i\cdot w_1(X_i; \alpha^*)\right] = \mathbb{E}[e(X_i; \alpha^*)w_1(X_i; \alpha^*)] = \mathbb{E}[h(X_i; \alpha^*)].
$

\textbf{Calculation of $B^*$.} 

$$B^* = \mathbb{E}[\phi^* \phi^{*\top}] = \begin{pmatrix}
  b_{11} & 0 & b_{13} \\
  0 & b_{22} & b_{23} \\
  b_{13}^\top & b_{23}^\top & b_{33}
\end{pmatrix},$$
where 
$$\begin{aligned}
  b_{11} &= \mathbb{E}\left[Z_i \cdot {\{Y_i(1) - \xi_1^*\}^2}\cdot {w_1(X_i; \alpha^*)^2}\right] = \mathbb{E}\left[{\{Y_i(1) - \xi_1^*\}^2}{e(X_i; \alpha^*)}w_1(X_i; \alpha^*)^2\right], \\
  b_{22} &= \mathbb{E}\left[(1 - Z_i) \cdot {\{Y_i(0) - \xi_0^*\}^2}\cdot {w_0(X_i; \alpha^*)^2}\right] \\
  &= \mathbb{E}\left[{\{Y_i(0) - \xi_0^*\}^2}\{1-e(X_i; \alpha^*)\}{w_0(X_i; \alpha^*)}^2\right].
\end{aligned}$$

\textbf{Calculation of $(A^*)^{-1}B^* (A^*)^{-\top}$.}

First, we notice that 
$$
A^* = \begin{pmatrix}
  a_{11} & 0 & a_{13} \\
  0 & a_{11} & a_{23} \\
  0 & 0 & a_{33}
\end{pmatrix} \Longrightarrow 
(A^*)^{-1} = \begin{pmatrix}
  a_{11}^{-1} & 0 & -a_{11}^{-1}a_{13}a_{33}^{-1}\\
  0 & a_{11}^{-1} & -a_{11}^{-1}a_{23}a_{33}^{-1}\\
  0 & 0 & a_{33}^{-1}
\end{pmatrix}
$$
Therefore, 
$$
\begin{aligned}
  &(A^*)^{-1}B^* (A^*)^{-\top} \\
  =& \begin{pmatrix}
    a_{11}^{-1} & 0 & -a_{11}^{-1}a_{13}a_{33}^{-1}\\
  0 & a_{11}^{-1} & -a_{11}^{-1}a_{23}a_{33}^{-1}\\
  0 & 0 & a_{33}^{-1}
  \end{pmatrix}\begin{pmatrix}
    b_{11} & 0 & b_{13} \\
    0 & b_{22} & b_{23} \\
    b_{13}^\top & b_{23}^\top & b_{33}
  \end{pmatrix}\begin{pmatrix}
    a_{11}^{-1} & 0 & 0\\
    0 & a_{11}^{-1} & 0\\
    -a_{33}^{-1}c_{13}^\top a_{11}^{-1} & -a_{33}^{-1}c_{23}^\top a_{11}^{-1} & a_{33}^{-1}
  \end{pmatrix} \\
  =& a_{11}^{-2}\begin{pmatrix}
    b_{11} - 2a_{13}a_{33}^{-1}b_{13}^\top + a_{13}a_{33}^{-1}a_{13}^\top  & -a_{13}a_{33}^{-1}b_{23}^\top - b_{13}a_{33}^{-1}a_{23}^\top + a_{13}a_{33}^{-1}a_{23}^\top & * \\
    -a_{13}a_{33}^{-1}b_{23}^\top - b_{13}a_{33}^{-1}a_{23}^\top + a_{13}a_{33}^{-1}a_{23}^\top & b_{22} - 2a_{23}a_{33}^{-1}b_{23}^\top + a_{23}a_{33}^{-1}a_{23}^\top & * \\
    * & * & *
  \end{pmatrix}.
\end{aligned}
$$
Hence, 
$$\begin{aligned}
  \mathrm{Var}(\hat{\tau}_w) &= \begin{pmatrix}
    1 & -1 & 0
  \end{pmatrix} (A^*)^{-1}B^* (A^*)^{-\top}\begin{pmatrix}
    1 \\ -1 \\ 0
  \end{pmatrix} \\
  &= a_{11}^{-2} \left[(b_{11} + b_{22} - 2(a_{13} - a_{23})a_{33}^{-1}(b_{13} - b_{23})^\top + (a_{13} - a_{23}) a_{33}^{-1} (a_{13} - a_{23})^\top)\right].
\end{aligned}$$

When the propensity scores are assumed known, $e'(X_i;\alpha^*) = 0$, and hence $a_{13} = a_{23} = 0$. Consequently, the variance reduces to $a_{11}^{-2}(b_{11} +b_{22})$. However, unlike ATE, there is no theoretical guarantee on whether assuming known propensity score leads to a larger or a smaller variance.

\subsection{Proof of Equation (18)} 

By the RMM model, $Y(z) = a W_e + z\tau + \epsilon$, where $\epsilon = W_e^\perp + \epsilon_{ori}$, $\bE[\epsilon]=\mu$, $\bV[\epsilon]=\sigma_y^2$, and $\epsilon_{ori}$ is the error term. Then, 
\begin{align}
    \xi_z = \frac{\mathbb{E}[h(X)Y(z)]}{\mathbb{E}[h(X)]} & = a\frac{\mathbb{E}[h(W_e)W_e]}{\mathbb{E}[h(W_e)]} + \frac{\mathbb{E}[h(W_e)z\tau]}{\mathbb{E}[h(W_e)]} + \frac{\mathbb{E}[h(W_e)\epsilon]}{\mathbb{E}[h(W_e)]} = a\zeta + z\tau + \mu,
\end{align}
where $\zeta = \mathbb{E}[h(W_e)W_e]/\mathbb{E}[h(W_e)]$, and the last equation holds because of Lemma~\ref{lem:rmm} (2).
Therefore, by plugging in $\xi_z$,
\begin{align*}
    &\mathbb{E}\left[\frac{(Y_i(1) - \xi_1)^2}{e(X_i)}h(X_i)^2\right] \nonumber \\
    =& \mathbb{E}\left[\left\{a\left(W_e - \zeta \right) + \epsilon - \mu \right\}^2 h(W_e)^2 \expit(W_e)^{-1} \right] \\
    =& \mathbb{E}\left[\mathbb{E}\left[\left\{a\left(W_e - \zeta \right) + \epsilon - \mu \right\}^2 h(W_e)^2 \expit(W_e)^{-1} \biggm| W_e \right]\right] \\
    =& \mathbb{E}\left[\left\{a^2\left(W_e - \zeta \right)^2 + \sigma_y^2\right\} h(W_e)^2 \expit(W_e)^{-1} \right] \\
    =& a^2 \mathbb{E} \left[\left(W_e - \zeta \right)^2 h(W_e)^2 \expit(W_e)^{-1} \right] + \sigma_y^2\mathbb{E}\left[ h(W_e)^2 \expit(W_e)^{-1} \right].
\end{align*}
The third equation holds because $\bV[\epsilon|W_e]=\bV[\epsilon]$ and the cross term equals zero, see Appendix~\ref{app:proof-thm3.2}. Similar derivation yields the other equation on $Y_i(0)$:
\begin{equation*}
    \mathbb{E}\left[\frac{(Y_i(0) - \xi_0)^2}{1-e(X_i)}h(X_i)^2\right] = a^2 \mathbb{E}\left[\left(W_e - \zeta \right)^2 \frac{h(W_e)^2}{1-\expit(W_e)} \right] + \sigma_y^2\mathbb{E}\left[ \frac{h(W_e)^2}{1-\expit(W_e)} \right].
\end{equation*}
Further define $g_z(W_e)={h(W_e)^2}[z\expit(W_e)+(1-z)\{1-\expit(W_e)\}]^{-1}$, and
\begin{align*}
    \mathcal{A}_z(h)= \bE[(W_e-\zeta)^2g_z(W_e)], \quad \mathcal{B}_z(h)= \bE[g_z(W_e)],
\end{align*}
then $V_w$ can be written as
\begin{equation}
    V_w = \frac{1}{\bE[h(W_e)]^2} \left[ a^2 \{\mathcal{A}_1(h)+ \mathcal{A}_0(h)\} + \sigma_y^2 \{\mathcal{B}_1(h)+ \mathcal{B}_0(h)\} \right].
\end{equation}
Plugging in $a^2=\rho^2S^2/\sigma_e^2$ and $\sigma_y^2=(1-\rho^2)S^2$ leads to
\begin{align*}
    V_w = & \frac{1}{\bE[h(W_e)]^2} \left[ \frac{\rho^2S^2}{\sigma_e^2} \{\mathcal{A}_1(h)+ \mathcal{A}_0(h)\} + (1-\rho^2)S^2 \{\mathcal{B}_1(h)+ \mathcal{B}_0(h)\} \right] \\
    & = \frac{S^2}{\bE[h(W_e)]^2} \left[ \frac{\rho^2}{\sigma_e^2} \{\mathcal{A}_1(h)+ \mathcal{A}_0(h)\} + (1-\rho^2) \{\mathcal{B}_1(h)+ \mathcal{B}_0(h)\} \right].
\end{align*}
Therefore, $V_w=S^2\widetilde{V}_w$, where
\begin{equation}
    \widetilde{V}_w = \frac{1}{\bE[h(W_e)]^2} \left[ \frac{\rho^2}{\sigma_e^2} \{\mathcal{A}_1(h)+ \mathcal{A}_0(h)\} + (1-\rho^2) \{\mathcal{B}_1(h)+ \mathcal{B}_0(h)\} \right]
\end{equation}
is the variance corresponding to the standardized effect $\widetilde{\tau}=\tau/S$.

\vspace{12pt}
\section{Examples}

\subsection{Empirical examples of normality of linear combination of covariates} 

\begin{figure}[htbp]
    \centering
    \includegraphics[width=\linewidth]{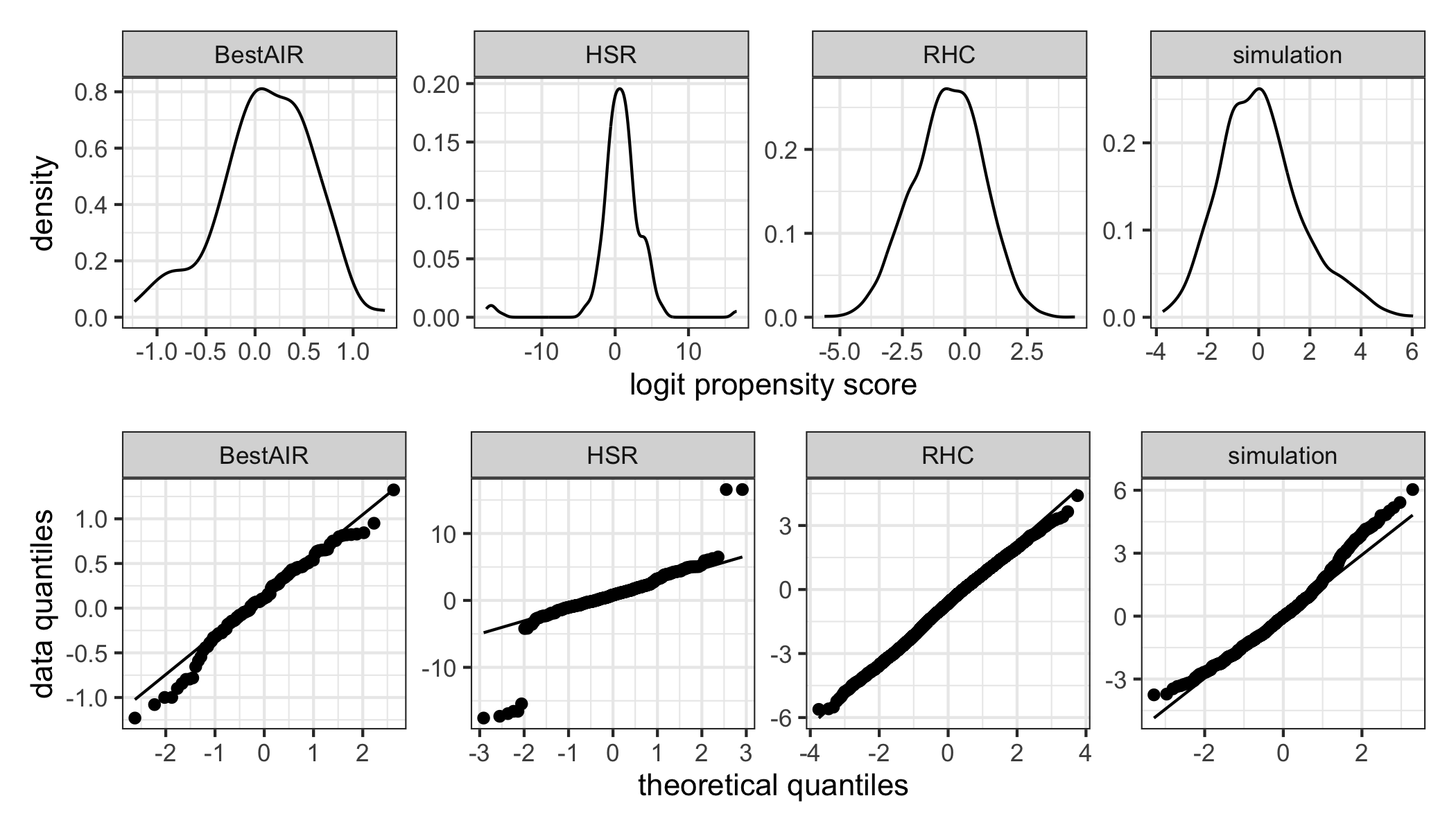}
    \caption{Density and Q-Q plots of the fitted logit propensity scores, a linear combination of covariates. The proximity of points to the reference line in the Q-Q plot shows the closeness of the distribution to a normal distribution.}
    \label{fig:normality}
\end{figure}

We consider the simulated data in Section 5 and three real world datasets: (i) RHC; (ii) the Best Apnea Interventions for Research (BestAIR) trial, a individually randomized trial with 169 units, with 9 continuous or binary covariates; more details can be found in \cite{zeng2021propensity}; (iii) HSR: an observational study on the racial disparities of breast cancer screening, with 56,480 units and 17 binary covariates; more details can be found in \cite{li2013propensity}. 
For each dataset, we fit a simple logistic model to estimate the propensity score $e(X) = \expit(\beta_0 + X'\beta)$, and we provide the density plot and Q-Q plot in Figure \ref{fig:normality}. Normality holds approximately in all the cases. 

\subsection{Examples of the Beta approximation of logit-normal distribution.} 
\begin{figure}
    \centering
    \includegraphics[width=\linewidth]{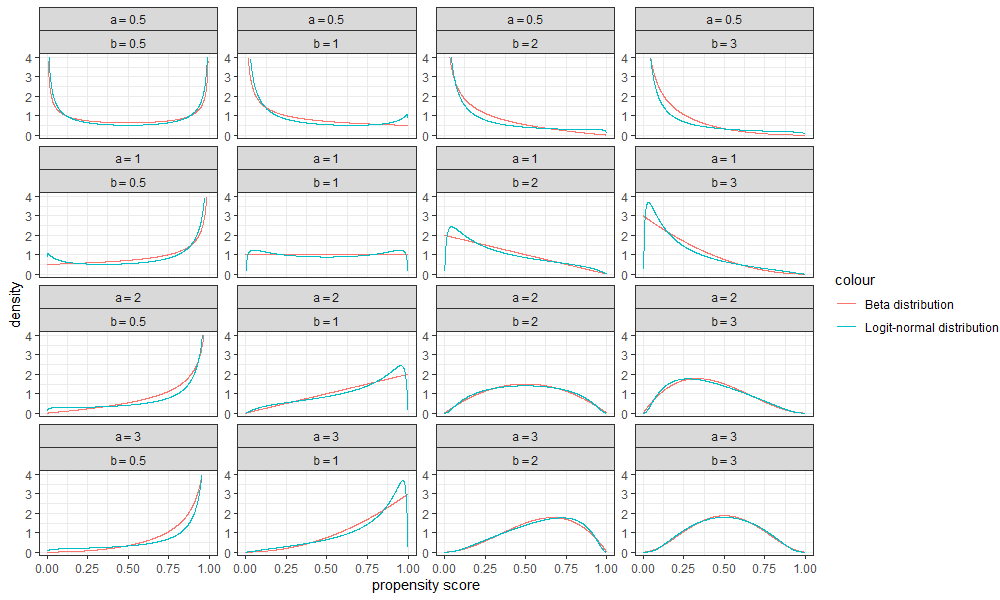}
    \caption{Examples of the Beta approximation to logit-normal distribution}
    \label{fig:beta_approx}
\end{figure}

Figure \ref{fig:beta_approx} presents the density of the Beta distribution over a grid of $(a, b)$ values within $\{0.5, 1, 2, 3\}$, and the density of the corresponding approximated logit-normal distributions.

\subsection{Calculation of overlap coefficient $\phi$ with fitted propensity scores} 

In practice, researchers may not have access to data and thus cannot calculate $\phi$ with fitted propensity scores. Nevertheless, we provide a computation method for interested readers. This method is more robust than directly estimating the density and can be useful for simulation studies.

By Bayes' theorem, 
\begin{equation}
    f_z(u) = \Pr(e(X) = u \mid Z = z) = \frac{\Pr(Z = z \mid e(X) = u) \Pr(e(X) = u)}{\Pr(Z = z)}.
\end{equation}
Hence, 
\begin{equation}
    \phi = \int_0^1\sqrt{f_0(u)f_1(u)}\,\mathrm{d} u = \frac{\int_0^1 \sqrt{u(1 - u)} \Pr(e(X) = u) \,\mathrm{d} u}{\sqrt{r(1 - r)}}.
\end{equation}
Let $F(u) = \Pr(e(X) \leq u)$ be the CDF of $e(X)$. Let $\widehat{e}_1, \widehat{e}_2, \dots, \widehat{e}_n$ be the estimated propensity scores of the units, and let $\widehat{F}(u) = \frac{1}{n}\sum_{i=1}^n \mathbb{I}(\widehat{e}_i \leq u)$ be the empirical CDF of $e(X)$. Then,
\begin{equation}
    \phi = \frac{\int_0^1 \sqrt{u(1 - u)} \,\mathrm{d} F(u)}{\sqrt{r(1 - r)}} \approx \frac{\int_0^1 \sqrt{u(1 - u)} \,\mathrm{d} \widehat{F}(u)}{\sqrt{r(1 - r)}} = \frac{1}{n}\cdot \frac{\sum_{i=1}^n \sqrt{\widehat{e}_i (1 - \widehat{e}_i)}}{\sqrt{r(1 - r)}}.
\end{equation}

\vspace{12pt}
\section{Additional Simulation Results}

\subsection{Distribution of propensity scores in two comparison groups of the simulated data in Section 5.1} 

\begin{figure}
    \centering
    \includegraphics[width=0.7\linewidth]{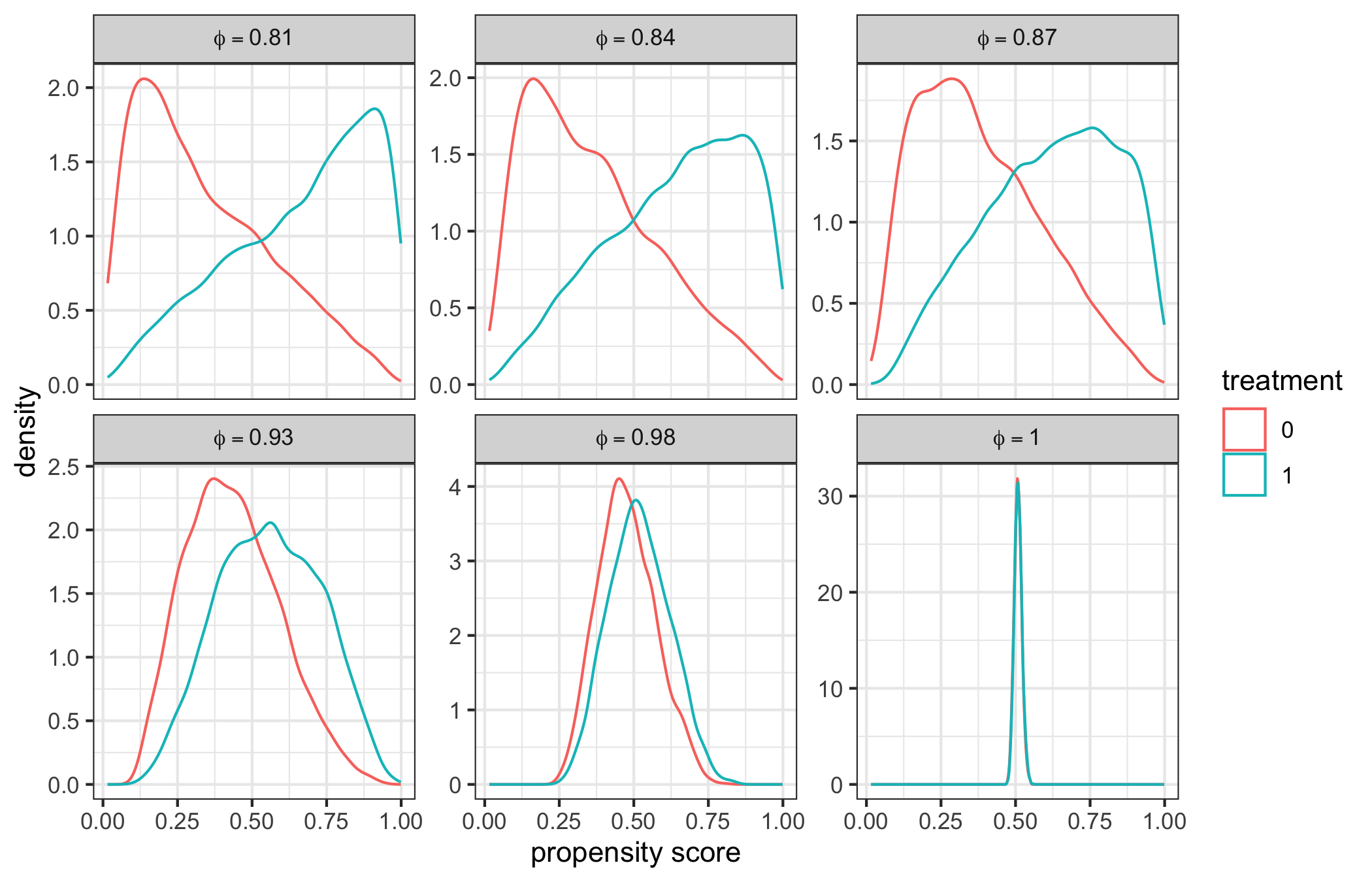}
    \caption{Distribution of the propensity scores in two comparison groups with fixed $r=0.5$ and various overlap coefficient $\phi=1.00, 0.98, 0.93, 0.87, 0.84, 0.81$, respectively, in simulation of synthetic data in Section 5.}
    \label{fig:sim1-ps}
\end{figure}

Figure \ref{fig:sim1-ps} visualizes the propensity score distributions in the two comparison groups under the data generation in Section 5.1 for each overlap coefficient.

\subsection{Simulations with binary outcomes} 

We repeat the simulation design in Section 5.1, and dichotomize the potential outcome $Y_i(z)$ into a binary variable $Y^b_i(z) = \mathbb{I}(Y_i(z) > -2)$. The observed outcome $Y^b_i = Z_iY^b_i(1) + (1 - Z_i)Y^b_i(0)$. The true ATE on $Y^b$ is $\tau^b = \mathbb{E}[Y^b_i(1) - Y^b_i(0)] = 0.089$. Assume that we estimate $\tau^b$ using the H\'{a}jek estimator with $Y^b_i$. Dichotomizing $Y_i(z)$ does not affect the propensity scores, and hence $\phi$ remain unchanged from Simulation 5.1, while $\rho^2, R^2$ changed. The pooled $S^2=r(1-r)=0.25$. We inversely solve the sample size to reach a nominal power of 0.80, and show the results in Table \ref{tab:sim2-result}.

\begin{table}
    \centering
    \caption{Sample size calculation by the proposed method, the two-sample $z$-test and the method in \citep{shook2022power} to achieve nominal power of 0.80 under various degrees of overlap. The power is obtained from random sampling of the calculated size from the $N = 1,000,000$ units. The Monte Carlo error for a power of 0.8 is 0.004. The treatment proportion $r$ is fixed at 0.5.}
    \label{tab:sim2-result}
    \begin{tabular}{cccccccccccc}
    \toprule
      &&& \multicolumn{3}{c}{Proposed} & \multicolumn{3}{c}{$z$-test} & \multicolumn{3}{c}{Shook-sa \& Hudgens} \\
 \cmidrule(lr){4-6} \cmidrule(lr){7-9} \cmidrule(lr){10-12}
      $\phi$ & $\rho^2$& $R^2$& Size & True PS & Est PS & Size & True PS & Est PS & Size & True PS & Est PS \\ \midrule
      1.00 &0.00&0.13& 992 & 0.81 & 0.85 & 977 & 0.80 & 0.83 & 977 & 0.80 & 0.83 \\
      0.98 &0.02&0.13& 1019 & 0.80 & 0.85 & 977 & 0.80 & 0.82 & 1016 & 0.80 & 0.85 \\
      0.93 &0.02&0.12& 1166 & 0.80 & 0.85 & 975 & 0.74 & 0.78 & 1157 & 0.80 & 0.85 \\
      0.88 &0.01&0.12& 1528 & 0.83 & 0.85 & 973 & 0.62 & 0.68 & 1489 & 0.80 & 0.85 \\
      0.84 &0.01&0.12& 1972 & 0.83 & 0.87 & 972 & 0.56 & 0.59 & 1914 & 0.82 & 0.86 \\
      0.81 &0.01&0.12& 2460 & 0.86 & 0.89 & 972 & 0.49 & 0.51 & 2342 & 0.83 & 0.87 \\ 
      \bottomrule
    \end{tabular}
\end{table}

From Table~\ref{tab:sim2-result}, the proposed method yields sample sizes achieving powers near 0.80, performing similarly to \cite{shook2022power}. In contrast, the two-sample $z$-test continuously leads to vastly underestimated sample size as the degree of overlap decreases.

\subsection{{Synthetic simulations with heterogeneous treatment effect and ATT and ATO estimand}}

Table~\ref{tab:hte-result-ATT-ATO} extends the HTE simulation to the ATT and ATO estimands, using the true estimand-specific effect size $\widetilde\tau = \tau^h/S$. The ATO remains well-calibrated, with empirical power near or above the nominal 0.80, consistent with its known efficiency and stability. The ATT, however, exhibits substantial power loss as overlap worsens, a sharp departure from the homogeneous case in Section 5.1 Table 2.

\begin{table}[ht]
    \centering
    \caption{Sample size (size) calculated by the proposed method to achieve the nominal power 0.8 under various degrees of overlap and HTE. The power is obtained from bootstrapping 10,000 random samples of the calculated size.}
    \label{tab:hte-result-ATT-ATO}
    \begin{tabular}{ccccccccccc}
    \toprule
      & \multicolumn{2}{c}{ATT} & \multicolumn{2}{c}{Power} & &\multicolumn{2}{c}{ATO} & \multicolumn{2}{c}{Power} \\ \cmidrule(lr){2-3} \cmidrule(lr){4-5} \cmidrule(lr){7-8} \cmidrule(lr){9-10}
      $\phi$ & $\tau^h$ & Size & True PS & Est PS & & $\tau^h$ & Size & True PS & Est PS \\ \midrule
      \multicolumn{10}{c}{\textit{covariates-driven HTE}} \\
      1.00 & 1.00 & 667 & 0.81 & 0.89 & & 1.00 & 665 & 0.80 & 0.89 \\
      0.98 & 1.03 & 675 & 0.78 & 0.86 & & 1.00 & 691 & 0.80 & 0.89 \\
      0.93 & 1.06 & 817 & 0.76 & 0.82 & & 0.99 & 761 & 0.81 & 0.88 \\
      0.87 & 1.08 & 1206 & 0.73 & 0.77 & & 0.99 & 864 & 0.81 & 0.89 \\
      0.83 & 1.09 & 1692 & 0.68 & 0.71 & & 0.99 & 938 & 0.83 & 0.90 \\
      0.81 & 1.09 & 2237 & 0.59 & 0.68 & & 0.99 & 993 & 0.82 & 0.89 \\ \midrule
      \multicolumn{10}{c}{\textit{propensity score-driven HTE}} \\
      0.98 & 1.19 & 473 & 0.80 & 0.84 & & 0.99 & 654 & 0.80 & 0.86 \\
      0.93 & 1.34 & 468 & 0.75 & 0.79 & & 0.97 & 744 & 0.80 & 0.87 \\
      0.87 & 1.45 & 602 & 0.71 & 0.73 & & 0.95 & 868 & 0.80 & 0.87 \\
      0.83 & 1.50 & 793 & 0.70 & 0.70 & & 0.95 & 952 & 0.81 & 0.87 \\
      0.81 & 1.53 & 1012 & 0.63 & 0.66 & & 0.94 & 1013 & 0.81 & 0.87 \\
    \end{tabular}
\end{table}

{As noted in Appendix~C.2, the sample size for the ATT is not guaranteed to be conservative; under HTE, this issue is amplified. Since HTE inflates $\bV\{Y(1)\}$ while leaving $\bV\{Y(0)\}$ unchanged, and all observed outcomes in the ATT's target population are $Y_i(1)$, the variance underestimation from the omitted $\bV\{\tau(X)\}$ term (Remark~1) is particularly consequential for the ATT. Practitioners targeting the ATT should exercise greater caution when effect heterogeneity is suspected.}

\subsection{RHC simulation with ATT and ATO estimands} 

The H\'{a}jek estimate of the ATT is 0.070 (standardized effect size 0.15). The sample size calculated to achieve power 0.8 is 2449.
With a sample of 2449 units, the bootstrap estimate of the power with $B = 1,000$ replications is 0.695 (95\% CI: 0.666 to 0.724). Unlike the ATE where we overestimate the sample size, our estimate for the ATT sample size is not guaranteed conservative. The H\'{a}jek estimate of the ATO is 0.075 (standardized effect size 0.16). The sample size calculated to achieve power 0.8 is 1546. With a sample of 1546 units, the bootstrap estimate of the power with $B = 1,000$ replications is 0.816 (95\% CI: 0.792 to 0.840).

\end{appendix}

\end{document}